\documentclass[showpacs,amsmath,amssymb,twocolumn,prl,superscriptaddress,notitlepage]{revtex4-1}

\usepackage{amsmath,amssymb,amsthm,mathrsfs,amsfonts,dsfont}
\usepackage{amsmath,bm}
\usepackage{bm}
\usepackage{braket}
\usepackage{color}
\usepackage{comment}
\usepackage{graphicx}

\newcounter{mynum}

\begin{document}

\title{Quantum-Enhanced Heat Engine Based on Superabsorption}

\author{Shunsuke Kamimura}
\email{s2130043@s.tsukuba.ac.jp}
\affiliation{Research Center for Emerging Computing Technologies,  National  Institute  of  Advanced  Industrial  Science  and  Technology  (AIST),1-1-1  Umezono,  Tsukuba,  Ibaraki  305-8568,  Japan.}
\affiliation{Faculty of Pure and Applied Sciences, University of Tsukuba, Tsukuba 305-8571, Japan}

\author{Hideaki Hakoshima}
\affiliation{Research Center for Emerging Computing Technologies,  National  Institute  of  Advanced  Industrial  Science  and  Technology  (AIST),1-1-1  Umezono,  Tsukuba,  Ibaraki  305-8568,  Japan.}
\affiliation{Center for Quantum Information and Quantum Biology, Osaka University, 1-3 Machikaneyama,Toyonaka, Osaka 560-8531, Japan}

\author{Yuichiro Matsuzaki}
\email{matsuzaki.yuichiro@aist.go.jp}
\affiliation{Research Center for Emerging Computing Technologies,  National  Institute  of  Advanced  Industrial  Science  and  Technology  (AIST),1-1-1  Umezono,  Tsukuba,  Ibaraki  305-8568,  Japan.}

\author{Kyo Yoshida}
\affiliation{Faculty of Pure and Applied Sciences, University of Tsukuba, Tsukuba 305-8571, Japan}

\author{Yasuhiro Tokura}
\email{tokura.yasuhiro.ft@u.tsukuba.ac.jp}
\affiliation{Faculty of Pure and Applied Sciences, University of Tsukuba, Tsukuba 305-8571, Japan}
\affiliation{Tsukuba Research Center for Energy Materials Science (TREMS), Tsukuba 305-8571, Japan}

\begin{abstract}

We propose a quantum-enhanced heat engine with entanglement.
The key feature of our scheme is superabsorption,
which facilitates enhanced energy absorption by entangled qubits.
Whereas a conventional engine with $N$ separable qubits provides power with a scaling of $P = \Theta (N)$,
our engine uses superabsorption to provide power with a quantum scaling of $P = \Theta(N^2)$.
This quantum heat engine also exhibits a scaling advantage over classical ones composed of $N$-particle Langevin systems.
Our work elucidates the quantum properties allowing for the enhancement of performance.

\end{abstract}

\maketitle

Quantum properties such as entanglement are important to realize desirable performance of devices.
Quantifying the performance of quantum devices often requires investigation of how the performance scales with the number of qubits $N$.
For example,
a quantum computer can solve certain problems exponentially faster than the best known classical algorithm \cite{shor1994algorithms,grover1997quantum,harrow2009quantum},
where the size of the problem corresponds to the number of qubits.
In quantum sensing,
the uncertainty in the target parameter scales as $\Theta (N^{-0.5})$ using separable qubits and $\Theta (N^{-1})$ using entangled qubits \cite{huelga1997improvement,giovannetti2011advances,degen2017quantum}.

Since the Industrial Revolution,
the properties and performances of heat engines have been successfully described
using a long-standing framework called thermodynamics \cite{carnot1872reflexions, callen1998thermodynamics}.
In previous decades,
thermodynamics has been generalized to classical small systems far from equilibria \cite{seifert2012stochastic},
and these systems cannot be understood without the information point of view \cite{landauer1961irreversibility, sagawa2008second, parrondo2015thermodynamics}.
This framework is referred to as stochastic thermodynamics.
It provides tighter constraints on the properties of systems than conventional thermodynamics,
such as fluctuation theorems \cite{jarzynski1997nonequilibrium, crooks1999entropy, hatano2001steady, seifert2005entropy}
and trade-off relations \cite{shiraishi2016universal, pietzonka2018universal},
and it is applicable to various research topics such as chemical reactions \cite{sekimoto2010stochastic, rao2016nonequilibrium}
and biological systems \cite{jarzynski2008thermodynamics, andrieux2008nonequilibrium}.

The development of microfabrication techniques has allowed devices to acquire quantum characteristics,
thereby facilitating various information processes that are much more efficient than conventional strategies,
as mentioned above.
Thus, there is a rapidly growing demand to establish thermodynamics generalized to quantum systems \cite{vinjanampathy2016quantum,deffner2019quantum}.
In this framework,
which is called
quantum thermodynamics~\cite{binder2018thermodynamics},
open quantum systems are considered as working media \cite{alicki1979quantum, quan2007quantum},
and relevant quantities such as work and heat
are defined by analogy with classical systems \cite{talkner2007fluctuation}.
Although the generalization of thermodynamics to quantum systems seems straightforward,
it significantly extends the scope of thermodynamics not only to heat engines composed of nanodevices \cite{bergenfeldt2014hybrid, zhang2014quantum, pekola2015towards, altintas2015rabi, peterson2019experimental}
but also to biological systems such as photosynthesis systems \cite{dorfman2013photosynthetic},
as quantum systems can provide a richer set of possibilities.
In particular,
the quantum version of the trade-off relation between power and efficiency is described by a measure of quantum coherence,
which has no counterpart in classical physics \cite{tajima2021superconducting}.

In the quantum thermodynamics, one of the main issues is
defining scaling advantages of quantum heat engines over classical ones \cite{tajima2021superconducting,hardal2015superradiant, niedenzu2018cooperative, kloc2019collective,watanabe2020quantum}.
In Ref.~\cite{tajima2021superconducting},
Tajima and Funo use an abstract system that has only two energies,
and there are $N_d$ degenerate states at each energy.
They show that quantum coherence among degenerate states can enhance the scaling
of a power with the number of degeneracy while the value of an efficiency is fixed.
This demonstrates the scaling enhancement of quantum engines at a finite temperature 
where the degeneracy of the system seems to play a role in defining the enhancement.

However, the model of Ref.~\cite{tajima2021superconducting}
is so abstract that we cannot easily understand the physics and mechanism behind the quantum enhancement.
Moreover, due to such nature of their model, it is not straightforward to find a physical counterpart of their model.
For a better interpretation of the quantum phenomena,
it is preferable to seek another concrete model
that provides the quantum enhancement with a physically realizable setting.

Here, we propose a quantum heat engine with a qubit-based model
that is relevant and applicable to many quantum information protocols.
The key feature of our model is a collective quantum phenomenon called superabsorption,
which allows for efficient energy exchange between the qubits and environment~\cite{higgins2014superabsorption, mirzaei2015superabsorption, brown2019light, yang2021realization}.
By using a physically realizable system,
we show how the enhanced power scaling of $\Theta (N^2)$
and the fixed value of the efficiency can be achieved with $N$ entangled qubits at the same time,
whereas a conventional engine with $N$ separable qubits provides a power with a scaling of $\Theta (N)$.
Our engine also beats classical engines composed of $N$ particles
obeying a Langevin equation where the power scales as $\Theta (N)$~\cite{shiraishi2016universal,shiraishi2019fundamental}.
Considering the microscopic model of the heat engine,
we add the understanding of quantum-enhanced performance.
More specifically, our results reveal that the description of the enhancement by the degeneracy is not generic;
rather the origin of the enhancement can be understood by a {\it connectivity} between quantum states where the system dynamics takes place.

{\it{Superabsorption.}}\textemdash
Superabsorption \cite{higgins2014superabsorption,brown2019light,mirzaei2015superabsorption,yang2021realization}
is the reverse process of superradiance.
In superradiance,
a collective emission is observed in an $N$-qubit system near the middle of the Dicke ladder \cite{dicke1954coherence,rehler1971superradiance,gross1982superradiance,brandes2005coherent}.
However,
an energy emission process is more dominant than an energy absorption process when a system is coupled with a white-noise environment;
thus,
it is not straightforward to observe superabsorption in a natural environment.
To overcome this limitation,
quantum control techniques can be used to enhance the absorption process.
In particular,
an interacting qubit system is coupled with a controlled environment for transition rate engineering;
then,
superabsorption can be achieved where the target two states chosen from the middle of the Dicke ladder have an enhanced transition \cite{higgins2014superabsorption}.

We introduce a Hamiltonian for superabsorption.
In this system,
we assume that the environment can be controlled by reservoir engineering.
In particular,
we consider a case in which the qubits are coupled with a leaky cavity \cite{fano1961effects,koshino2005quantum},
and this coupling induces an energy relaxation on the qubits with a Lorentzian form factor \cite{purcell1946spontaneous,bienfait2016controlling}.
The Hamiltonian $\hat{H}_{\text{tot}}$ of the total system is given by
\begin{align}
	\hat{H}_{\text{tot}} &= \hat{H}_{N} + \hat{H}_{\text{E}} + \hat{H}_{\text{int}}, \\
 	\hat{H}_{N} &= \omega_A \hat{J}_z + \Omega \hat{J}_z^2 , \nonumber \\
 	\hat{H}_{\text{E}} &= \int_{-\infty}^{\infty} dk \ \omega_k \hat{B}_k^{\dagger} \hat{B}_k , \ \ \ \omega_k = |k|, \nonumber \\
 	\hat{H}_{\text{int}} &= \int_{-\infty}^{\infty} dk  \left( \hat{J}_+ + \hat{J}_-  \right) \left( \xi (\omega_k) \hat{B}_k + \xi^* (\omega_k) \hat{B}^{\dagger}_k \right) , \nonumber \\
 			\xi (\omega) &= \sqrt{ \frac{ \Delta \omega }{ 2 \pi } } \frac{ g }{ \omega - \omega_c - i \Delta \omega /2},
\end{align}
where $\hat{H}_{N}$ ($\hat{H}_{\text{E}}$) denotes a Hamiltonian for the $N$-qubit system
(environment)
and $\hat{H}_{\text{int}}$ denotes an interaction Hamiltonian between the system and
environment.
In our system,
all $N$ qubits have the same
frequency $\omega_A$ and interact with each other in an all-to-all manner with strength $\Omega$.
For the engineered bosonic environment,
a mode with wave number $k$
(and energy $\omega_k = |k|$)
is collectively coupled to the $N$-qubit system with
the complex function
$\xi (\omega_k)$.
The cavity,
with frequency $\omega_c$,
is coupled to the $N$-qubit system with strength $g$,
and $\Delta \omega$ is the decay rate of the cavity.
The bosonic operators $\hat{B}_k$ satisfy the commutation relations $[ \hat{B}_k, \hat{B}^{\dagger}_{k'} ] = \delta (k - k ')$.
The collective operators $\hat{J}_z$ and $\hat{J}_{\pm}$ are defined by summations of the Pauli operators for all the qubits as
$\hat{J}_z = \frac{1}{2} \sum_{i=1}^N \hat{\sigma}_z^{(i)}$
and
$\hat{J}_{\pm} = \sum_{i=1}^N \hat{\sigma}_{\pm}^{(i)}$,
where $\hat{\sigma}_z=|\text{e}\rangle \langle \text{e}|-|\text{g}\rangle \langle \text{g}|$ , $\hat{\sigma}_+ = |\text{e}\rangle \langle \text{g}|$, $ \hat{\sigma}_- =  (\hat{\sigma}_+)^{\dagger}$
and $\ket{ \text{e} }$ ($\ket{ \text{g} }$) denotes the excited (ground) state of a qubit.
By introducing another collective operator $\hat{J}^2$ that represents the total angular momentum,
we define $\ket{J,M}$ as a Dicke state,
which is a simultaneous eigenstate of the operators $\hat{J}^2$ and $\hat{J}_z$
with eigenvalues $J (J+1)$ and $M$,
respectively \cite{dicke1954coherence}.

When the initial state of the system belongs to a subspace spanned by the Dicke states
$\ket{M} = \left| \frac{N}{2}, M \right\rangle$
having the maximum total angular momentum,
the dynamics under consideration is totally confined in the same subspace.
This subspace is called the Dicke ladder,
within which the Hamiltonian $\hat{H}_{N}$ is diagonal as
\begin{align}
	\hat{H}_{N} &= \sum_{M} E_M \ket{M} \! \bra{M} , \ E_M = \omega_A M + \Omega M^2.
\end{align}
In addition,
we define an energy difference
$\Delta_M = E_M - E_{M-1} = \omega_A + (2M-1) \Omega $
and a transition frequency
$\omega_M = |\Delta_M| $
between the Dicke states $\ket{M}$ and $\ket{M-1}$
$ \left( M = - \frac{N}{2} , - \frac{N}{2} +1, \ldots , \frac{N}{2} \right) $.

By adopting the standard Born--Markov and rotating-wave approximation,
for an $N$-qubit quantum state $\hat{\rho}_S$ that is diagonal in the Dicke 
states
$\{ \ket{M} \}$,
we can derive the following Gorini--Kossakowski--Sudarshan--Lindblad (GKSL) master equation \cite{higgins2014superabsorption}:
\begin{align}
	\frac{d \hat{\rho}_S}{dt}  = \sum_M a_M \left( \Gamma_M^{\downarrow} \mathcal{D} [ \hat{L}_M^{\downarrow} ] [ \hat{\rho}_S ] + \Gamma_M^{\uparrow} \mathcal{D} [ \hat{L}_M^{\uparrow} ] [ \hat{\rho}_S ] \right).
	\label{eq:GKSL_SA_main}
\end{align}
For a positive $\Delta_M$,
the factor $\Gamma_M^{\downarrow} = \kappa_M ( 1 + n_M )$
($\Gamma_M^{\uparrow} = \kappa_M n_M $)
is a transition coefficient for the dynamics
$\ket{M} \mapsto \ket{M-1}$
($\ket{M-1} \mapsto \ket{M}$),
where $n_M = 1/ \left( e^{\beta \omega_M} - 1 \right)$ is defined as the Bose--Einstein occupation number with an inverse temperature $\beta$
and $\kappa_M = 4 \pi | \xi (\omega_M) |^2$ is the value of the
spectral density at frequency $\omega_M$.
The dynamics $\ket{M} \mapsto \ket{M-1}$ ($\ket{M-1} \mapsto \ket{M}$) 
is induced by a Lindblad operator
$\hat{L}_M^{\downarrow} =|M-1\rangle \langle M|$ $\left( \hat{L}_M^{\uparrow} = ( \hat{L}_M^{\downarrow} )^{\dagger} \right)$.
(For a negative $\Delta_M$, the definitions of the coefficients and the Lindblad operators are, respectively, interchanged.)
In a two-level system defined as $\mathcal{H}_{M} = \{ \ket{M}, \ket{M-1} \}$,
a detailed-balance condition
$\Gamma_M^{\downarrow} / \Gamma_M^{\uparrow} = e^{ \beta \Delta_M }$
is satisfied for each $M$.
The superoperator
$\mathcal{D}$ is given by $\mathcal{D} [ \hat{L} ] [ \hat{\rho} ] = \hat{L} \hat{\rho} \hat{L}^{\dagger} - \frac{1}{2} ( \hat{L}^{\dagger} \hat{L} \hat{\rho} + \hat{\rho} \hat{L}^{\dagger} \hat{L} )$
for arbitrary operators $\hat{L}$ and $\hat{\rho}$,
and $a_M = \left( \frac{N}{2} + M \right) \left( \frac{N}{2} - M + 1 \right)$.
In particular,
$a_M$ quantifies the enhancement of the transition rate of $\mathcal{H}_M$.
Specifically,
for an odd number $N$ (as assumed throughout this Letter),
the label $M = 1/2$ gives us the largest factor $a_{1/2} = (N+1)^2/4$.
Thus,
we obtain the maximum enhancement of the transition rate within a subspace
$\mathcal{H}_{1/2} = \{ \ket{1/2} , \ket{-1/2} \}$,
which we call the {\it effective two-level system} (E2LS).
It is worth noting that $\ket{1/2}$ and $\ket{-1/2}$ are highly entangled states used for many other applications in quantum information processing \cite{dicke1954coherence,rehler1971superradiance,campbell2009characterizing,hyllus2010not,toth2012multipartite,wu2017generation,kasture2018scalable,hakoshima2020efficient,kasai2021anonymous}.

The key aspect of superabsorption is to confine the dynamics within the E2LS.
Such a confinement can be realized by setting $\Omega \gg \Delta \omega$
and $\omega_c = \omega_{1/2}$ (which we adopt throughout this Letter).
In this case,
owing to the frequency selectivity,
the environment is strongly coupled only with the E2LS,
and the energy absorption transition $\ket{-1/2} \mapsto \ket{1/2}$ becomes much more relevant than the energy emission process $\ket{-1/2} \mapsto \ket{-3/2}$.

\begin{figure}
	\begin{center}
		\includegraphics[clip,width=8.5cm,bb=0 0 775 450]{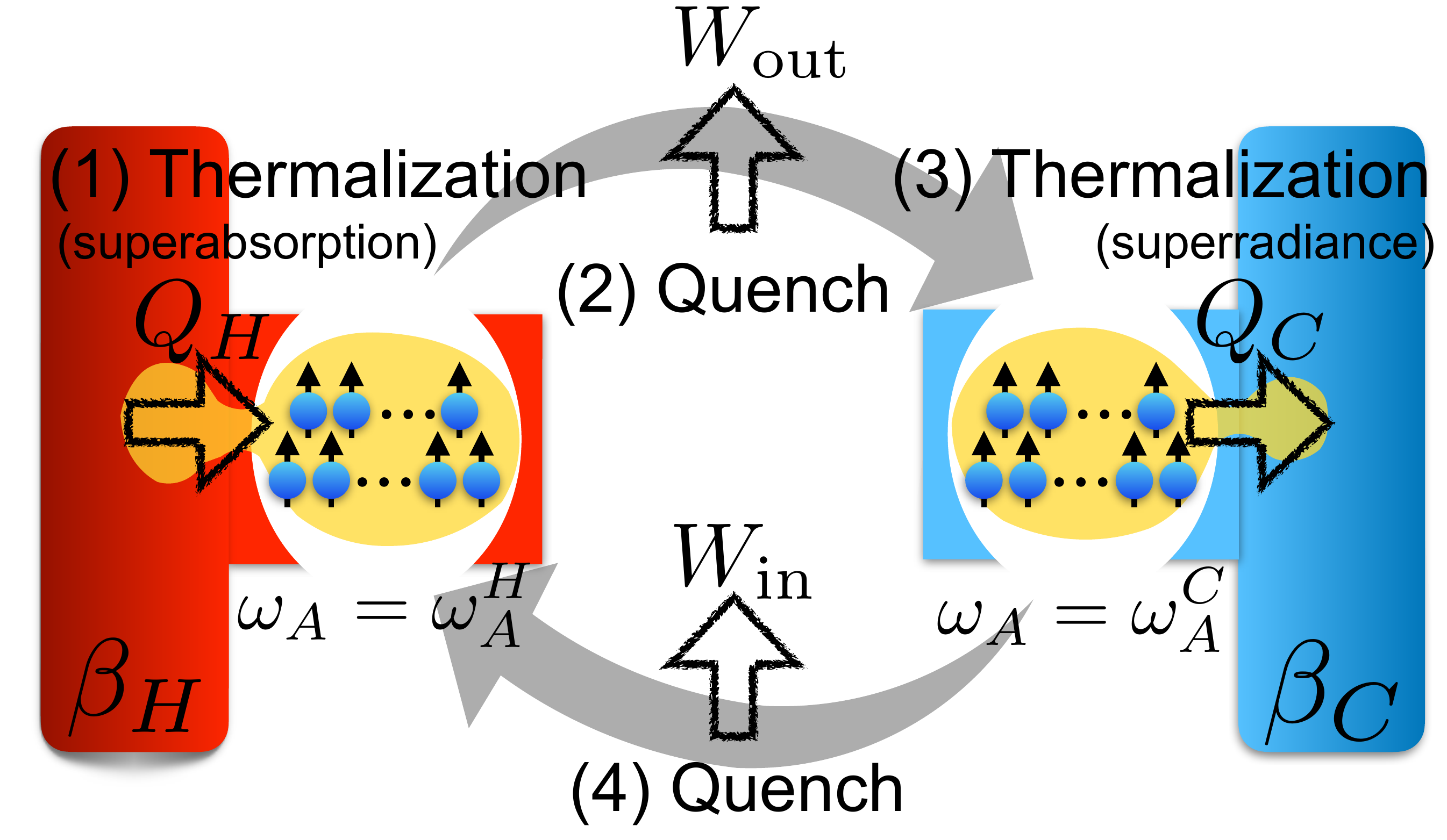}
		\caption{Schematic of protocol for heat engine based on superabsorption.
		$N$ qubits interact with a cavity coupled with a thermal bath.
		This configuration is useful for tailoring the properties of the environment for the qubits.
		By repeated thermalization of the qubits and quenching of the qubit frequency,
		we can extract the work.
		A heat engine with separable states corresponds to a case with $N=1$,
		and we can operate $N$ separable engines in parallel.}
		\label{fig:SAQHE_schematic}
	\end{center}
\end{figure}

{\it{Heat engine based on superabsorption.}}\textemdash
Here, we describe our protocol of the quantum-enhanced heat engine based on superabsorption
(see Fig. \ref{fig:SAQHE_schematic}).
The $N$-qubit system is a working medium in our scheme.
We employ a high-temperature bath and a low-temperature bath.
The inverse temperature and cavity frequency of the high(low)-temperature bath are
$\beta _H$($\beta _C$) and $\omega _c^H$($\omega _c^C$), respectively.
The initial state $\hat{\rho}_S (0)$ is
diagonal in the Dicke states
$\{ \ket{M} \}_M$ as $\hat{\rho}_S (0) = \sum_M p_M (0) \ket{M} \! \bra{M}$.
A cycle of the heat engine consists of four strokes as follows:

\noindent
{\it Stroke 1: Thermalization with a high-temperature bath.}
The $N$-qubit system is coupled with a high-temperature bath $\beta_H$ for a period $\tau_H$.
During this process, the qubits are resonant with the cavity as $\omega_c^H = \omega_A^H$,
and the energy of the Dicke state $\ket{M}$ is given by
$E_M^H = M \omega_A^H + M^2 \Omega $.
The dynamics of this thermalization stroke from
$\hat{\rho}_S (0)$ to $\hat{\rho}_S (\tau_H) = \sum_M p_M (\tau_H) \ket{M} \! \bra{M}$
is governed by Eq.~(\ref{eq:GKSL_SA_main}),
and the change in the energy of the $N$-qubit system is interpreted as a heat input $Q_H$ to the system:
$Q_H = \sum_M E_M^H \left[ p_M (\tau_H) - p_M (0) \right]$.

\noindent
{\it Stroke 2: Quenching of the qubit frequency $\omega_A^H \mapsto \omega_A^C$.}
After decoupling all the qubits from the heat bath $\beta_H$,
the qubit frequency is simultaneously changed from $\omega_A^H$ to $\omega_A^C$.
We assume that this is performed in a much shorter time than the time required for a single cycle of our engine.
As the instantaneous Hamiltonian $\hat{H}_{N}$ always commutes with $\hat{\rho}_S (\tau_H)$,
the state remains in $\hat{\rho}_S (\tau_H)$.
For this stroke, the energy change of the system is interpreted as a work output $W_{\text{out}}$ from the system:
$W_{\text{out}} = \sum_M \left( E_M^H - E_M^C \right) p_M (\tau_H)$.
Here, $E_M^C = M \omega_A^C + M^2 \Omega $ is the energy of $\ket{M}$ after this quenching stroke.

\noindent
{\it Stroke 3: Thermalization with a low-temperature bath.}
The $N$-qubit system is coupled with a low-temperature bath $\beta_C$ for a period $\tau_C$,
and a resonant condition $\omega_c^C = \omega_A^C$ is assumed.
The dynamics in this stroke is described by
$\hat{\rho}_S (\tau_H) \mapsto \hat{\rho}_S (\tau) = \sum_M p_M (\tau) \ket{M} \! \bra{M}$
($\tau = \tau_H + \tau_C$).
Similar to Stroke 1,
a heat output $Q_C$ to the bath is defined as
$Q_C = \sum_M E_M^C \left[ p_M (\tau_H) - p_M (\tau) \right]$.

\noindent
{\it Stroke 4: Quenching of the qubit frequency $\omega_A^C \mapsto \omega_A^H$.}
After decoupling all the qubits from the heat bath $\beta_C$,
the qubit frequency is simultaneously changed from $\omega_A^C$ to $\omega_A^H$.
Again, we assume that we can ignore the quenching time for this stroke.
For the same reason as that in the case of Stroke 2,
the quantum state remains in $\hat{\rho}_S (\tau)$ during this stroke,
and a work input $W_{\text{in}}$ to the system is defined as
$W_{\text{in}} = \sum_M \left( E_M^H - E_M^C \right) p_M (\tau)$.

We define the efficiency $\eta$ and power output $P$ of the heat engine cycle as
\begin{align}
	\eta &= \frac{ W_{\text{ext}} }{Q_H} ,
	& P &= \frac{ W_{\text{ext}} }{ \tau },
\end{align}
where $W_{\text{ext}} = W_{\text{out}} - W_{\text{in}}$ denotes the extractable work.
We introduce an efficiency deficit as $\Delta \eta = \eta_C - \eta$,
where $\eta_C = 1-\beta_H/\beta_C$ is the Carnot efficiency.


Now, we explain the choice of parameters.
We set the thermalization periods as
$\tau_H = \epsilon /( a_{1/2} \Gamma_{1/2}^{H \downarrow} ) $
and
$\tau_C = \epsilon / ( a_{1/2} \Gamma_{1/2}^{C \downarrow} )$,
where $\epsilon$ denotes a positive dimensionless constant.
Then,
the cycle period $\tau = \tau_H + \tau_C$
scales as $\tau = \Theta (N^{-2})$.
For the subsequent analytical discussion,
$\epsilon$ should be much smaller than 1 so that higher-order terms
$O(\epsilon^2)$ can be ignored.
Moreover,
we choose the initial state
$\hat{\rho}_S (0) = \sum_M p_M (0) \ket{M} \! \bra{M} $
as
\begin{align}
	p_{-1/2} (0) &= \frac{ 2 }{ 2 + e^{ - \beta_H \omega_A^H } + e^{ - \beta_C \omega_A^C } },
	\label{initialone}
\end{align}
where $p_{1/2} (0) = 1 - p_{-1/2} (0) $
and $p_M (0) = 0$ for $M \neq \frac{1}{2} , - \frac{1}{2} $.
This choice is for an analytical form of the power and efficiency,
as will be described later.
Now, we consider a ratio $\chi_{\text{conf}}$
between $ p_{1/2} (\tau_H) - p_{1/2} (0) $ and $ p_{-3/2} (\tau_H) - p_{-3/2} (0) $,
which quantifies the population change in the E2LS compared with that in $\ket{-3/2}$.
It is worth noting that,
because we consider a low-temperature condition
($\beta_H \omega_A^H , \beta_C \omega_A^C \gg 1 $),
the population of $\ket{3/2}$ is negligible.
According to this analysis,
we choose parameters that satisfy the following condition:
\begin{align}
	\chi_{ \text{conf} } = \left[ 1 + 16 \left( \frac{ \Omega }{ \Delta \omega} \right)^2 \right] \frac{ e^{ - \beta_H \omega_A^H } - e^{ - \beta_C \omega_A^C } }{2} \gg 1.
	\label{confinecondition}
\end{align}
Thus, as long as this condition (\ref{confinecondition}) is satisfied,
the dynamics of the system is nearly confined in the E2LS.
The condition (\ref{confinecondition}) implies that both the reservoir engineering and the coupling between qubits are essential,
because we cannot satisfy this condition with either a white-noise environment ($\Delta \omega \to \infty$) or noninteracting qubits ($\Omega = 0$).
Therefore,
the concept of superabsorption is crucial for realizing the proposed engine.

Here, we consider a simplified scenario in which the dynamics is perfectly confined in the E2LS to calculate the power and efficiency,
although we will discuss more realistic cases with a finite leakage from the E2LS later.
When we adopt the initial state described by Eq.~(\ref{initialone}) and $p_{1/2} (0) = 1 - p_{-1/2} (0)$,
by disregarding both the leakage
and the higher-order terms $O(\epsilon^2)$,
we can construct a closed trajectory of the quantum state after a heat engine cycle;
the populations of $\ket{1/2}$ and $\ket{-1/2}$ do not change after the cycle.
Then,
we obtain the following forms of the efficiency deficit and power, respectively:
\begin{align}
	\Delta \eta_{ \text{E2LS} } &= \frac{ \omega_A^C }{ \omega_A^H } - \frac{ \beta_H }{ \beta_C } \ge 0 , \\
	P_{\text{E2LS}} &= a_{1/2} P_{N=1} ,
	\label{pscalingq}
\end{align}
where the power output $P_{N=1}$ for a 1-qubit system is explicitly given by
\begin{align}
	P_{N = 1} = \frac{ \gamma_P \left( e^{ - \beta_H \omega_A^H } - e^{ - \beta_C \omega_A^C } \right)}{ 4 - \left( e^{ - \beta_H \omega_A^H } + e^{ - \beta_C \omega_A^C } \right)^2 } \left( \omega_A^H - \omega_A^C \right).
\end{align}
Here,
$\gamma_P = 8 g^2 / \Delta \omega $ represents a modified relaxation rate owing to the Purcell effect \cite{goy1983observation,bienfait2016controlling,purcell1946spontaneous}.
From Eq.~(\ref{pscalingq}),
because we have $a_{1/2} = \frac{1}{4} ( N + 1 )^2$,
we obtain the quantum-enhanced performance $P = \Theta (N^2)$ at a finite temperature,
which is significantly different from the performance $P_{\rm{sep}} = \Theta (N)$ obtained with $N$ separable qubits.

Here, we consider what properties of quantum systems contribute to the scaling advantage of performance.
The Dicke state $\ket{1/2}$ ($\ket{-1/2}$) with $N$ qubits is
an equal-weight superposition of all computational bases with 
$\frac{N+1}{2}$ $\left( \frac{N-1}{2} \right)$ qubits in $| \text{e} \rangle $
and $\frac{N-1}{2}$ $\left( \frac{N+1}{2} \right)$ qubits in $| \text{g} \rangle $,
and the energies corresponding to $\ket{1/2}$ and $\ket{-1/2}$ both have
an exponentially large degeneracy given by 
${}_{N} C_{{ (N+1) }/2} \sim 2^{N}/\sqrt{N}$.
However, by applying the jump operator of Eq.~(\ref{eq:GKSL_SA_main}),
which flips only one spin,
to a computational basis of $\ket{1/2}$ ($\ket{-1/2}$),
we have only $\frac{N+1}{2}$
bases of $\ket{-1/2}$ ($\ket{1/2}$), and we define this value as a \textit{connectivity}.
This value provides us a matrix element of the operator $\hat{J}_{\pm}$ between $\ket{1/2}$ and $\ket{-1/2}$,
and its square gives the resulting scaling factor of the transition rate $a_{1/2} = \frac{1}{4} (N+1)^2$ in our system.
This shows that, for the quantum enhancement, the degeneracy is not generic;
rather the {\it connectivity} of the quantum states
between one degenerated subspace and another
induced by the system-environment interaction is crucial.
Moreover, our results provide a unified understanding of both our model and that of Tajima and Funo
\cite{tajima2021superconducting}.
In their case, the number of connectivity coincides with that of degeneracy,
and thus
our results lead to a conclusion that their scaling advantage also comes from the number of connectivity~\cite{caption}.


Now, we consider the dependence of the confinement performance $\chi_{\text{conf}}$ on the efficiency deficit $\Delta \eta_{\text{E2LS}}$ in our engine.
In particular, we consider the case in which we tune only $\omega_A^C$ to change $\Delta \eta_{\text{E2LS}}$.
Then, $\chi_{\text{conf}}$ can be rewritten as
\begin{align}
	\chi_{\text{conf}} &= \frac{ \beta_C \omega_A^H  }{ 2 e^{ \beta_H \omega_A^H } } \left[ 1 + 16 \left( \frac{\Omega}{\Delta \omega} \right)^2 \right] \Delta \eta_{\text{E2LS}} + O (  \Delta \eta_{\text{E2LS}}^2 ).
\end{align}
This implies that $\chi_{\text{conf}}$ is linearly dependent on $\Delta \eta_{\text{E2LS}}$.
When we take the Carnot limit $\Delta \eta_{\text{E2LS}} \to 0$ by tuning $\omega_A^C$ and fixing the other parameters,
$\chi_{\text{conf}}$ approaches zero,
and the system is no longer confined in the E2LS.
Meanwhile,
for a fixed $\Delta \eta_{\text{E2LS}}$,
by choosing a larger value of $\Omega/\Delta \omega$,
we can maintain the confinement condition of $\chi_{\text{conf}}\gg 1$.
Thus,
controlling the parameter $\Omega/\Delta \omega$ via reservoir engineering is crucial for our scheme.

Next,
we investigate a trade-off relation between the power and the efficiency of our engine.
Tajima and Funo derived a trade-off relation $P/\Delta \eta \le \mathcal{B}_{\text{TF}} $
for quantum heat engines described by GKSL master equations,
where the upper bound $ \mathcal{B}_{\text{TF}} $ quantifies a (time-averaged) measure of quantum coherence during the heat engine cycle~\cite{tajima2021superconducting}.
To the best of our knowledge,
our study is the first to apply the general formula to an $N$-qubit system,
and we prove that $\mathcal{B}_{\text{TF}}= \Theta (N^2)$.
Thus,
our heat engine scheme attains this upper bound in terms of
the scaling with $N$.
For a classical model of heat bath described by a Langevin equation,
there is a known bound of
$P/ \Delta \eta \le \mathcal{B}_{\text{SST}} = \Theta (N)$
for an $N$-particle system,
as discussed in Refs.~\cite{shiraishi2016universal,shiraishi2019fundamental}.
Therefore, the trade-off performance $P/\Delta \eta = \Theta (N^2)$ of our heat engine reflects the scaling advantage over such classical engines as well~\cite{caption}.

\begin{figure}
	\begin{center}
		\includegraphics[clip,width=9cm,bb=0 0 1175 725]{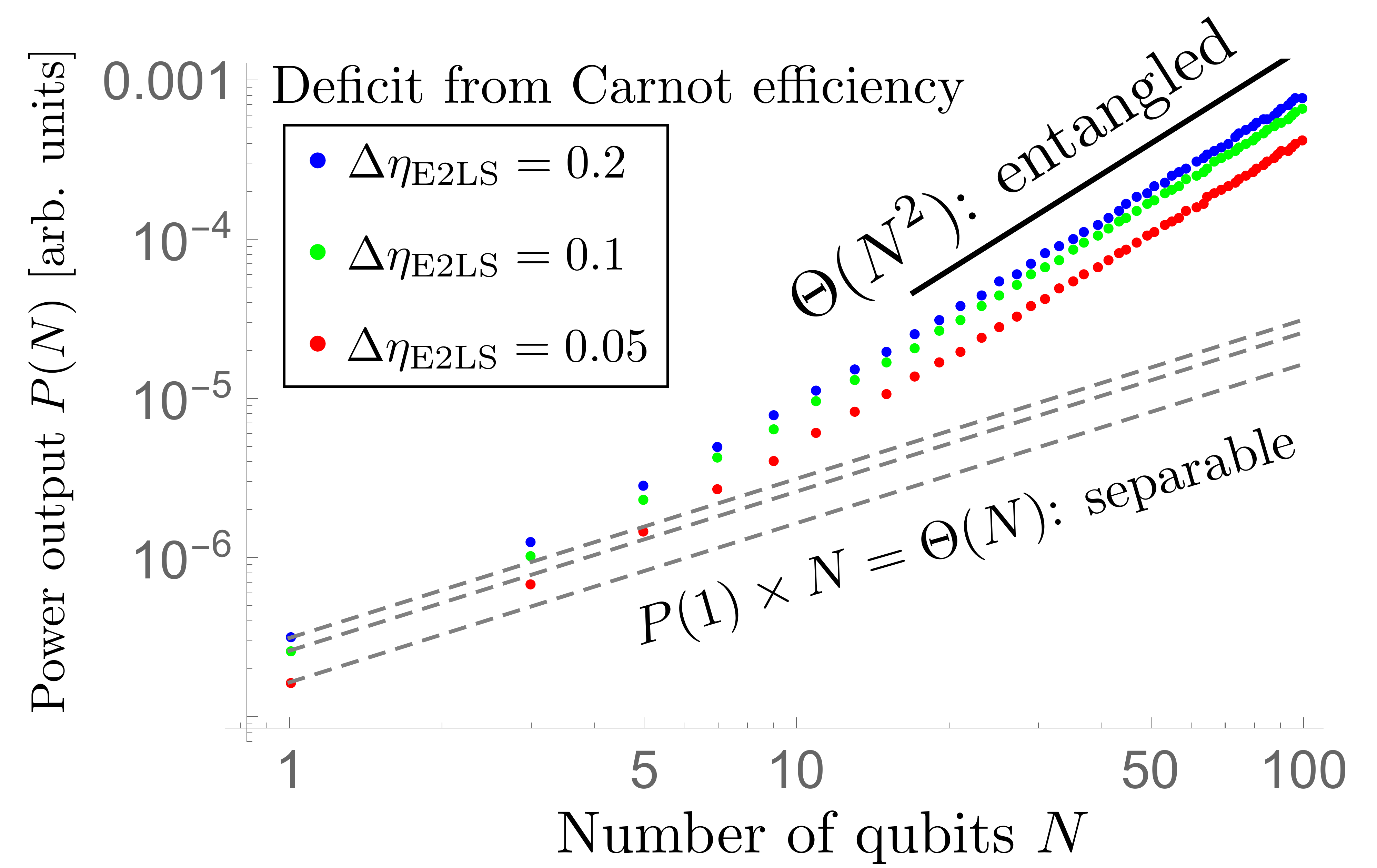}
		\caption{Dependency of the power output $P (N)$ on the number of qubits $N$.
		The numerical values of the parameters are chosen as
		$\omega_A^H/2\pi = 1$ GHz,
		$\Omega/2 \pi = 31$ MHz,
		$T_H = (k_B \beta_H)^{-1} = 20$ mK,
		$T_C = (k_B \beta_C)^{-1} = 10$ mK,
		$g/ 2\pi = 10$ kHz,
		$\Delta \omega / 2\pi = 1$ MHz,
		and $\epsilon = 0.001$.
		To change the efficiency as
		$\Delta \eta_{\text{E2LS}} = 0.2, 0.1, 0.05$,
		we adjust $\omega_A^C$ as
		$\omega_A^C / 2\pi  = 0.7$ GHz, $0.6$ GHz, $0.55$ GHz,
		and accordingly,
		$\chi_{\text{conf}}$ changes as
		$\chi_{\text{conf}} = 399, 248, 139$, respectively \cite{caption}.}
		\label{fig:SAQHE_power_Nsc}
	\end{center}
\end{figure}

{\it{Numerical results.}}\textemdash
Here, we present numerical results on the performance of our heat engine under the effect of a finite leakage from the E2LS.
First,
we estimate the number of cycles $n_{\text{conf}}$ during which the quantum state is significantly confined to the E2LS as
$ n_{\text{conf}} = \frac{1}{ a_{-1/2} \left( \Gamma_{-1/2}^{H \downarrow} \tau_H +  \Gamma_{-1/2}^{C \downarrow} \tau_C \right) }\simeq  \left[ 1 + 16 \left( \Omega / \Delta \omega \right)^2 \right] / \left( 2 \epsilon \right) $.
For our settings,
$n_{\text{conf}} \simeq 7.69 \times 10^6$.
Second,
for $\Delta \eta_{\text{E2LS}} = 0.05$ and $N = 31$,
we numerically calculate $ \Delta \eta$ and $P$ for $5000 \ (\ll n_{\text{conf}} )$ heat engine cycles,
and we find that the results deviate from
$ \Delta \eta_{\text{E2LS}}$ and $P_{\text{E2LS}}$
only by a few percent at most, respectively
(the detailed settings are given in the caption of Fig. \ref{fig:SAQHE_power_Nsc}).
From these results,
we conclude that the confinement in the E2LS is sufficiently strong to claim that we approximately have a closed trajectory of the quantum state after each cycle.
Finally, we numerically calculate the power output $P(N)$ against $N$ with several values of $\omega_A^C$.
As we fix the other parameters,
the change in $\omega_A^C$ induces the change in $\Delta \eta_{\text{E2LS}}$.
We plot the power outputs against the number of qubits, as shown in Fig. \ref{fig:SAQHE_power_Nsc}.
Here, we define $P(N)$ as the power output of the first cycle for each $N$.
As $\Delta \eta_{\text{E2LS}}$ decreases,
the power $P$ decreases for the fixed $N$.
However, importantly, the behavior $P = \Theta (N^2)$ is observed regardless of the value of $\Delta \eta_{\text{E2LS}}$.
Therefore,
for a large number of qubits,
we can achieve both high power and high efficiency in our engine with entanglement,
compared with the conventional engine with separable states.

{\it{Conclusion.}}\textemdash
We proposed a quantum-enhanced heat engine with a power output that exhibits a quantum scaling with the number of qubits
at a finite temperature.
Our engine is fueled by an entanglement-enhanced energy absorption process called superabsorption,
where the dynamics of an $N$-qubit system is approximately confined in a subspace spanned by two highly entangled states.
We analytically showed that, as long as the confinement is significant,
our engine achieves a power of $P= \Theta (N^2)$ with $N$ entangled qubits,
whereas a conventional engine with $N$ separable qubits provides a power of $P= \Theta (N)$.
Moreover,
we numerically observed the same scaling advantage even under the effect of a finite leakage to the other states.
We elucidate the mechanism of quantum enhancement of performance,
and show that a connectivity of the quantum states between one degenerated subspace and another 
induced by the system-environment interaction plays an important role
in achieving the scaling advantage with the quantum heat engine.
Our proposal is also important for realizing next-generation quantum devices such as high-performance refrigerators for quantum systems
\cite{kosloff2014quantum, brunner2014entanglement, tan2017quantum}.

We thank H. Tajima and K. Funo for their insightful feedback.
We also thank an anonymous reviewer for invaluable comments regarding the importance of connectivity.
This work was supported by MEXT's Leading Initiative for Excellent Young Researchers and JST PRESTO (Grant No. JPMJPR1919), Japan.
Y. T. acknowledges support from JSPS KAKENHI (No. 20H01827) and JST's Moonshot R$\&$D (Grant No. JPMJMS2061).

{\it{Note added.}}\textemdash
Recently,
we became aware of a related study that uses a collective effect for a quantum refrigerator
\cite{kloc2021superradiant}.

\nocite{*}

\bibliographystyle{apsrev4-1}

%

\end{document}



\title{Supplementary Material: Quantum-enhanced heat engine based on superabsorption}

\author{Shunsuke Kamimura}
\email{s2130043@s.tsukuba.ac.jp}
\affiliation{Research Center for Emerging Computing Technologies,  National  Institute  of  Advanced  Industrial  Science  and  Technology  (AIST),1-1-1  Umezono,  Tsukuba,  Ibaraki  305-8568,  Japan.}
\affiliation{Faculty of Pure and Applied Sciences, University of Tsukuba, Tsukuba 305-8571, Japan}

\author{Hideaki Hakoshima}
\affiliation{Research Center for Emerging Computing Technologies,  National  Institute  of  Advanced  Industrial  Science  and  Technology  (AIST),1-1-1  Umezono,  Tsukuba,  Ibaraki  305-8568,  Japan.}
\affiliation{Center for Quantum Information and Quantum Biology, Osaka University, 1-3 Machikaneyama,Toyonaka, Osaka 560-8531, Japan}


\author{Yuichiro Matsuzaki}
\email{matsuzaki.yuichiro@aist.go.jp}
\affiliation{Research Center for Emerging Computing Technologies,  National  Institute  of  Advanced  Industrial  Science  and  Technology  (AIST),1-1-1  Umezono,  Tsukuba,  Ibaraki  305-8568,  Japan.}

\author{Kyo Yoshida}
\affiliation{Faculty of Pure and Applied Sciences, University of Tsukuba, Tsukuba 305-8571, Japan}

\author{Yasuhiro Tokura}
\email{tokura.yasuhiro.ft@u.tsukuba.ac.jp}
\affiliation{Faculty of Pure and Applied Sciences, University of Tsukuba, Tsukuba 305-8571, Japan}
\affiliation{Tsukuba Research Center for Energy Materials Science (TREMS), Tsukuba 305-8571, Japan}

\maketitle


This supplementary material consists of four sections.
The first section is devoted to briefly summarize a trade-off relation originally derived by Tajima and Funo \cite{tajima2021superconducting}.
In the second section, we introduce an $N$-qubit system exhibiting superabsorption,
and apply the trade-off relation to our heat engine by the superabsorption.
In the third section, we compare our system with a specific model considered by Tajima and Funo, and explain differences between them.
In the forth section, we discuss possible experimental realizations of our heat engine. 

We use the unit where $\hbar = c = k_B = 1$.

\section{Trade-off relation}

In this section, we discuss a trade-off relation of a quantum heat engine originally derived by Tajima and Funo \cite{tajima2021superconducting}.
We summarize the trade-off relation, and explicitly describe all of quantities and their definitions relevant to the relation.

\subsection{GKSL master equation}

In this subsection, we review a GKSL master equation of a quantum system
coupled with an environment, which we will use for the quantum heat engine.
Suppose that a quantum system is coupled with an environment,
and the system Hamiltonian is given by $\hat{H}$.
It is known that, under some assumptions such as a Born-Markov approximation and a rotating wave approximation,
the dynamics of a density operator $\hat{\rho}_S (t)$ of the system at time $t$
is described by the following GKSL master equation:
\begin{align}
	\frac{d}{dt} \hat{\rho}_S (t) &= -i [ \hat{H} , \hat{\rho}_S (t) ] + \mathcal{D} [ \hat{\rho}_S (t)], &
	& \text{where } \mathcal{D} [ \hat{\rho} ] \ceq \sum_{\omega} \gamma_{\omega} \left[ \hat{L}_{\omega} \hat{\rho} \hat{L}_\omega^{\dagger} - \frac{1}{2} \{ \hat{L}_{\omega}^{\dagger} \hat{L}_{\omega} , \hat{\rho} \}\right].
\end{align}
Here, the first term of the right-hand side of the first equation corresponds to
the unitary dynamics induced by the system Hamiltonian $\hat{H}$.
We assume that a Lamb shift is negligibly small.
The second term $\mathcal{D} [\hat{\rho}_S (t)]$ 
corresponds to the dissipative dynamics induced by the environment.
This is characterized by the dissipative coefficients $\gamma_{\omega} \ge 0$ and the corresponding Lindblad operators $\hat{L}_{\omega}$.
For a positive (negative) $\omega$, $\hat{L}_{\omega}$ is associated with an energy relaxation (thermal excitation) to decrease (increase) energy of $\omega$ with respect to $\hat{H}$,
while $\hat{L}_{\omega}$ is associated with dephasing for $\omega=0$. 
It is worth mentioning that the Hamiltonian $\hat{H}$ can be degenerate,
and actually this degeneracy plays a central role in an enhancement of a heat engine performance in quantum regime \cite{tajima2021superconducting}.

To deduce thermodynamic properties, following the work
by Tajima and Funo \cite{tajima2021superconducting}, we assume the three conditions below:
\begin{align}
	(1) \ \frac{\gamma_{\omega}}{\gamma_{- \omega}} &= e^{ \beta \omega }, &
	(2) \ [ \hat{L}_{\omega} , \hat{H} ] &= \omega \hat{L}_{\omega}, &
	(3) \ \hat{L}_{- \omega} = \hat{L}_{\omega}^{\dagger}.
\end{align}
The condition (1) means that the environment is a thermal bath whose inverse temperature is given by $\beta$.
This  is conventionally called detailed balance condition.
The last two conditions are satisfied in most cases where a GKSL master equation is derived from a microscopic model.

\subsection{Current-dissipation trade-off relation}

For a GKSL master equation system introduced previously,
a trade-off relation for a heat current $J (t)$ and an entropy production rate (dissipation) $\dot{\sigma} (t)$ at time $t$ can be derived.
This relation takes the following form: 
\begin{align}
	\frac{ J^2 (t) }{ \dot{\sigma} (t) } \le \mathcal{A} (t) .
\end{align}
Here, the upper bound $\mathcal{A} (t)$ has information about the system and its environment, 
and an explicit definition of $\mathcal{A} (t)$ is provided for classical \cite{shiraishi2016universal} and quantum master equation systems \cite{tajima2021superconducting}, respectively.
Here, the heat current $J(t)$ and the entropy production rate $\dot{\sigma} (t)$
for a quantum state $\hat{\rho} (t)$ are respectively defined as
\begin{align}
	J (t) & \ceq \text{Tr} \left[ \hat{H} \frac{d \hat{\rho} (t) }{dt} \right], &
	\dot{\sigma} (t) & \ceq \dot{S} (t) - \beta J (t),
\end{align}
where $S(t) = - \text{Tr} [ \hat{\rho} (t) \log \hat{\rho} (t) ] $ is
the von-Neumann entropy of the system,
and $\dot{S} (t)$ represents its time derivative.

In particular, for the GKSL master equation system under our consideration, 
the explicit form of $\mathcal{A}$ can be written as follows:
\begin{align}
	\mathcal{A} (t) &\ceq \frac{ A_{\text{cl}} (t) + A_{\text{qm}}  (t)}{2}, & 
	\text{where } A_{\text{cl}}  (t) &= \text{Tr} [ \hat{X} \hat{\rho}_{\text{sd}}  (t)] 
	\ \ \ \text{and} \ \ \ 
	A_{\text{qm}}  (t) = \mathcal{C}_{\hat{X}} \mathcal{C}_{l_1} (\hat{\rho}_{\text{bd}}  (t)).
\end{align}
Explicit definitions of these quantities are shown below.

First, we define the operator $\hat{X}$
as
\begin{align}
	\hat{X} \ceq \sum_{\omega} \omega^2 \gamma_{\omega} \hat{L}_{\omega}^{\dagger} \hat{L}_{\omega} \ \text{: positive operator}.
\end{align}
Next, we define $\hat{\rho}_{\text{bd}} (t)$ and $\hat{\rho}_{\text{sd}} (t)$.
We describe the Hamiltonian $\hat{H}$ as
\begin{align}
	\hat{H} = \sum_{e} \sum_{a=1}^{d_e } E_e \ket{e, a} \! \bra{e, a},
\end{align}
where $E_e$ denotes the energy and $\ket{e, a}$ denotes the energy eigenstate.
The label $e$ represents the energy $E_e$, and 
different labels specify different energies
(i.e. $e \neq e' \Rightarrow E_e \neq E_{e'}$).
The other label $a$ is for the degeneracy of the energy $E_e$,
and
$d_e$ is the number of the degenerate energy eigenstates $\{ \ket{e,a} \}_a$.
Then, we define a ``block diagonal (bd)'' state $\hat{\rho}_{\text{bd}} (t)$ for the quantum state $\hat{\rho} (t)$ as
\begin{align}
	\hat{\rho}_{\text{bd}} (t) &\ceq \sum_{e} \hat{\Pi}_e \hat{\rho} (t) \hat{\Pi}_e,
	\label{eq:bd_state}
\end{align}
where $\hat{\Pi}_e \ceq \sum_{a = 1}^{d_e} \ket{e,a} \! \bra{e,a}$ denotes a projection operator to the energy eigenspace.
Due to the projection operators, this state $\hat{\rho}_{\text{bd}} (t)$ does not have off-diagonal elements for any two eigenstates with different energies.
Similarly, we define the ``strictly diagonal (sd)'' state $\hat{\rho}_{\text{sd}} (t)$ as
\begin{align}
	\hat{\rho}_{\text{sd}} (t) &\ceq \sum_{e} \sum_{a=1}^{d_e} \hat{\Pi}_{e,a} \hat{\rho} (t) \hat{\Pi}_{e,a} ,
	\label{eq:sd_state}
\end{align}
where $\hat{\Pi}_{e,a}  \ceq  \ket{e,a} \! \bra{e,a}$.
This ``sd'' state $\hat{\rho}_{\text{sd}} (t)$ has no off-diagonal elements for all two 
orthogonal states regardless of their degeneracy.
By the definitions so far, we can calculate $A_{\text{cl}} (t) = \text{Tr} [ \hat{X} \hat{\rho}_{\text{sd}} (t) ]$, which is time-dependent through the ``sd'' state.
Finally, we address the definition of $\mathcal{C}_{\hat{X}}$ and $\mathcal{C}_{l_1}(\hat{\rho}_{\text{bd}})$.
The factor $\mathcal{C}_{\hat{X}}$ is defined as
\begin{align}
	\mathcal{C}_{\hat{X}} \ceq \max_{e,a,a': a \neq a'} | \! \bra{e, a} \hat{X} \ket{e, a'} \! |.
\end{align}
This factor is time independent, and is determined by the set of eigenstates $\ket{e, a}$ and the operator $\hat{X}$.
This factor evaluates the largest off-diagonal element of $\hat{X}$
among all degenerate energy eigenstates.
Meanwhile, the factor $\mathcal{C}_{l_1} (\hat{\rho}_{\text{bd}} (t) )$ is
given as follows:
\begin{align}
	\mathcal{C}_{l_1} (\hat{\rho}_{\text{bd}} (t) ) \ceq \sum_{(e, a) \neq (e' , a' ) } | \! \bra{e, a} \hat{\rho}_{\text{bd}} (t) \ket{e', a' } \! | .
\end{align}
This norm $\mathcal{C}_{l_1}$ is a well-known measure in the resource theory of coherence \cite{streltsov2017colloquium},
and this evaluates the amount of a coherence (stored in the state $\hat{\rho}_{\text{bd}} (t)$)
among all the energy eigenstates.

\subsection{Power-efficiency trade-off relation}

We consider a heat engine protocol to use two heat baths with inverse temperatures $\beta_H$ and $\beta_C \ ( > \beta_H )$.
When the system dynamics forms a closed trajectory in a single heat engine cycle, 
a trade-off relation for the power $P$ and the efficiency deficit $\Delta \eta = \eta_{\text{C}} - \eta$ can be derived
from the relation for $J (t)$ and $\dot{\sigma} (t)$ \cite{tajima2021superconducting}:
\begin{align}
	\frac{J^2 (t) }{\dot{\sigma} (t)} & \le \mathcal{A} (t) \ \ \ 
	\Rightarrow \ \ \ 
	\frac{P}{\Delta \eta } \le \alpha \overline{ \mathcal{A} }  \ \ \ \text{where } \alpha \ceq \frac{\beta_C \eta_C }{ ( 2 - \eta_C )^2} .
	\label{tfgeneral}
\end{align}
Here, $\overline{\mathcal{A}}$ represents a time average of the upper bound $\mathcal{A} (t)$ in the entire dynamics of the single heat engine cycle with a period $\tau$ i.e. $\overline{\mathcal{A}} \ceq \frac{1}{\tau} \int_0^{\tau} dt \mathcal{A} (t) $.


\section{Superabsorption and trade-off relation}
In this section, we first introduce an $N$-qubit system exhibiting superabsorption,
and we apply the trade-off relation to the system.
As a result, we find that a performance of our heat engine saturates the upper bound of the trade-off relation in terms of scaling with $N$.

\subsection{Superabsorption}
In this subsection, we introduce an $N$-qubit system exhibiting superabsorption.
As discussed in the main text, the Hamiltonian $\hat{H}_{\text{SA}}$ of the superabsorption 
$N$-qubit system is given by
\begin{align}
	\hat{H}_{\text{SA}} &\ceq \omega_A \hat{J}_z + \Omega \hat{J}_z^2,
\end{align}
where $\hat{J}_z \ceq \frac{1}{2} \sum_{i=1}^N \hat{\sigma}_z^{(i)}$ is a collective Pauli operator,
$\omega_A$ is a qubit frequency,
and $\Omega$ is a strength of the all-to-all interaction between the qubits.
In the superabsorption system introduced in the main text, the system dynamics is totally confined in a subspace
spanned by Dicke states with maximum total angular momentum.
Thus, we can omit the dynamics outside of this subspace,
and we can easily diagonalize $\hat{H}_{\text{SA}}$ within this subspace:
\begin{align}
	\hat{H}_{\text{SA}} \ceq \sum_{M= -N/2}^{N/2} E_M \ket{M} \! \bra{M},
\end{align}
where $E_M = \omega_A M + \Omega M^2$ is the eigenenergy of the eigenstate $\ket{M}$.
The eigenstate $\ket{M}$ is called Dicke state, and this is an eigenstate
of the collective Pauli operator $\hat{J}_z$ with an eigenvalue $M$ $\left( M = - \frac{N}{2} , - \frac{N}{2} + 1, \ldots , \frac{N}{2} \right)$.
For convenience, we here define an energy difference
$\Delta_M$ and a transition frequency $\omega_M$ of a 2-level system $\{ \ket{M} , \ket{M -1} \}$ as 
\begin{align}
	\Delta_M &\ceq E_M - E_{M-1} = \omega_A + (2M-1) \Omega, &
	\omega_M \ceq |\Delta_M|.
\end{align}
Throughout this supplementary material, we consider the case where $N$ is an odd number as we assume in the main text.

Applying a Born-Markov approximation and a rotating-wave approximation, we can obtain a GKSL master equation for a quantum state $\hat{\rho} (t)$ of the $N$-qubit system as follows:
\begin{align}
	\frac{d \hat{\rho} (t) }{dt} = -i [ \hat{H}_{\text{SA}} , \hat{\rho} (t) ] + \mathcal{D}_{\text{SA}} [ \hat{\rho} (t) ].
\end{align}
The dissipation term $\mathcal{D}_{\text{SA}} $ for the superabsorption is explicitly given as
\begin{align}
	\mathcal{D}_{\text{SA}} [\hat{\rho}] \ceq \sum_{M=-N/2}^{N/2} a_M \kappa (\omega_M) \left[ \left( 1 + n (\omega_M) \right) \mathcal{D} [\hat{L}_M ] [\hat{\rho} ] + n (\omega_M) \mathcal{D} [\hat{L}^{\dagger}_M ] [\hat{\rho} ] \right],
\end{align}
where $a_M = \left( \frac{N}{2} + M \right) \left( \frac{N}{2} - M + 1 \right)$ is a coefficient,
$\kappa (\omega) $ is a spectral density of the environment, and $n (\omega) = 1/ (e^{\beta \omega} - 1)$ is a Bose-Einstein function.
Here, the superoperator $\mathcal{D} $ is defined
as $\mathcal{D} [\hat{L} ] [\hat{\rho}] \ceq \hat{L} \hat{\rho} \hat{L}^{\dagger} - \frac{1}{2} \{ \hat{L}^{\dagger} \hat{L} , \hat{\rho} \} $,
and the
Lindblad operator $\hat{L}_M$ is defined as
\begin{align}
	\hat{L}_M \ceq \ket{M-1} \! \bra{M},
\end{align}
for $\Delta_M \ge 0$.
(Instead, when $\Delta_M < 0$, $L_M$ is defined as $\hat{L}_M \ceq \ket{M} \! \bra{M-1}$.)

\subsection{Upper bound $\mathcal{A} (t)$ for superabsorption}
In this subsection, by applying the trade-off relation in Eq.~\eqref{tfgeneral}
to the $N$-qubit system exhibiting superabsorption \cite{higgins2014superabsorption},
we calculate an upper-bound of $P/ \Delta \eta $ of our heat engine.
We assume that the quantum state $\hat{\rho} (t)$ of the system is diagonal with respect to the Dicke states $\ket{M}$:
\begin{align}
	\hat{\rho} (t) &= \sum_M p_M (t) \ket{M} \! \bra{M}, &
	0 \le p_M (t) &\le 1, &
	\sum_M p_M (t) &= 1.
\end{align}
Also, we assume that all energy difference $\Delta_M$ is positive:
\begin{align}
	\Delta_M &= \omega_A + (2M-1) \Omega \ge 0 , &
	M &= - \frac{N}{2}, - \frac{N}{2} + 1 , \ldots , \frac{N}{2} - 1,  \frac{N}{2}.
\end{align}
Then, the Lindblad operator can be explicitly written as $\hat{L}_M = \ket{M-1} \! \bra{M}$ for all $M$.

In this setup, we can calculate the upper bound $\mathcal{A} = ( A_{\text{cl}} + A_{\text{qm}} )/2$.
For $A_{\text{cl}} = \text{Tr} [ \hat{X} \hat{\rho}_{\text{sd}} ]$, the positive operator $\hat{X}$ is given by
\begin{align}
	\hat{X} =\sum_{M} \Delta_M^2 \gamma_M^{\uparrow} \ket{M} \! \bra{M} + \sum_M \Delta_M^2 \gamma_M^{\downarrow} \ket{M-1} \! \bra{M-1},
\end{align}
where the coefficients are respectively defined as
\begin{align}
	\gamma_M^{\uparrow} &= a_M \kappa (\omega_M) n (\omega_M) , & \gamma_M^{\downarrow} &= a_M \kappa (\omega_M) ( 1 + n (\omega_M) ).
\end{align}
The ``sd'' state $\hat{\rho}_{\text{sd}} (t) = \sum_{e,a} \hat{\Pi}_{e,a} \hat{\rho} (t) \hat{\Pi}_{e,a}$ is also simplified as
\begin{align}
	\hat{\rho}_{\text{sd}} (t) &= \sum_{M = -N/2}^{N/2} \frac{p_M (t)}{ {}_N C_{N/2 + M} } \sum_{\text{all configuration}}^{{}_N C_{N/2 + M}}
	\ket{ \underbrace{\text{e} \cdots \text{e}}_{N/2 + M} \underbrace{\text{g} \cdots \text{g}}_{N/2 - M} } \! \bra{ \underbrace{\text{e} \cdots \text{e} }_{N/2 + M} \underbrace{ \text{g} \cdots \text{g} }_{N/2 - M} },
	\label{kamimuraformulaone}
\end{align}
where $\sum_{\text{all configuration}}^{{}_N C_{N/2 + M}}$ denotes
a summation about all ${}_N C_{N/2 + M}$ distinct configurations of the labels $\{ \underbrace{\text{e} , \ldots , \text{e} }_{N/2+M} , \underbrace{\text{g} , \ldots , \text{g} }_{N/2-M} \} $.
For simplicity, we use the symbol $\sum_{\text{a.c.}}$ for this summation.
About the Dicke state $\ket{M}$, we can also derive the following:
\begin{align}
	\braket{ \underbrace{\text{e} , \ldots , \text{e} }_{N/2+M} \underbrace{\text{g} , \ldots , \text{g} }_{N/2-M} | M } &= \frac{1}{\sqrt{ {}_N C_{N/2 + M} } } &
	\text{ for all configuration.} \label{kamimuraformulatwo}
\end{align}
By using Eqs. \eqref{kamimuraformulaone} and \eqref{kamimuraformulatwo},
$A_{\text{cl}}$ can be evaluated as
\begin{align}
	A_{\text{cl}} (t) &= \sum_M \frac{p_M (t) }{ {}_N C_{N/2 + M} } \text{Tr} \left[ \hat{X} \sum_{\text{a.c.}}^{{}_N C_{N/2 + M}}
	\ket{ \underbrace{\text{e} \cdots \text{e}}_{N/2 + M} \underbrace{\text{g} \cdots \text{g} }_{N/2 - M} } \! \bra{ \underbrace{\text{e} \cdots \text{e}}_{N/2 + M} \underbrace{\text{g} \cdots \text{g}}_{N/2 - M} }  \right] \\
	&= \sum_M \frac{p_M (t) }{ {}_N C_{N/2+M} }
	\sum_{\text{a.c.}}^{{}_N C_{N/2 + M}} \bra{ \underbrace{\text{e} \cdots \text{e}}_{N/2 + M} \underbrace{\text{g} \cdots \text{g}}_{N/2 - M} } \hat{X} \ket{ \underbrace{\text{e} \cdots \text{e}}_{N/2 + M} \underbrace{\text{g} \cdots \text{g}}_{N/2 - M} } \\
	&=  \sum_M \frac{p_M (t) }{ {}_N C_{N/2+M} }
	\left[ \frac{ \Delta_M^2 \gamma_M^{\uparrow} }{ {}_N C_{N/2+M} } + \frac{ \Delta_{M+1}^2 \gamma_{M+1}^{\downarrow} }{ {}_N C_{N/2+M} }  \right] \sum_{\text{a.c.}}^{{}_N C_{N/2 + M}} 1 \\
	&=  \sum_M \frac{p_M (t) }{ {}_N C_{N/2+M} }
	\left[ \Delta_M^2 \gamma_M^{\uparrow} + \Delta_{M+1}^2 \gamma_{M+1}^{\downarrow} \right] .
\end{align}
For a heat engine based on superabsorption, most of the population is confined in the 2-level system $\{ \ket{1/2} , \ket{-1/2} \} $ i.e. $p_{1/2} (t) + p_{-1/2} (t) \simeq 1$.
In this case, $A_{\text{cl}} (t)$ can be approximately evaluated as
\begin{align}
	A_{\text{cl}} (t) \simeq \sum_{M=1/2, -1/2} \frac{p_M (t) }{ {}_N C_{N/2+M} }
	\left[ \Delta_M^2 \gamma_M^{\uparrow} + \Delta_{M+1}^2 \gamma_{M+1}^{\downarrow} \right] 
	\le \sum_{M=1/2, -1/2} \frac{ 1 }{ {}_N C_{N/2+M} }
	\left[ \Delta_M^2 \gamma_M^{\uparrow} + \Delta_{M+1}^2 \gamma_{M+1}^{\downarrow} \right] .
\end{align}
From the Stirling formula $n! \simeq \sqrt{2 \pi n } \left( n/e \right)^n$,
we roughly evaluate how $A_{\text{cl}} (t) $ scales with
$N$ as
\begin{align}
	A_{\text{cl}} (t) = \Theta \left( \frac{N^2}{2^N} \right).
\end{align}
To obtain this equation, we used $\gamma_{\pm 1/2}^{\downarrow (\uparrow)} = \Theta (N^2)$.
Therefore, for the heat engine based on
superabsorption, this term appearing in the upper bound of the trade-off relation exponentially decreases with $N$, and it's going to be tiny.

For the other term $A_{\text{qm}}$ appearing in the trade-off relation, the first remark is that the ``bd'' state of the Dicke state $\ket{M} \! \bra{M}$ is the same state.
Thus, 
\begin{align}
	\hat{\rho}_{\text{bd}} (t)
	= \sum_{e} \hat{\Pi}_e \hat{\rho} (t) \hat{\Pi}_e
	= \sum_{e,M} p_M (t) \hat{\Pi}_e \ket{M} \! \bra{M} \hat{\Pi}_e
	= \sum_M p_M (t) \ket{M} \! \bra{M}
	= \hat{\rho} (t).
\end{align}
We calculate the two factors $\mathcal{C}_{\hat{X}} $ and $\mathcal{C}_{l_1} (\hat{\rho}_{\text{bd}} (t) )$ included in $A_{\text{qm}} = \mathcal{C}_{\hat{X}} \mathcal{C}_{l_1} (\hat{\rho}_{\text{bd}} (t) )$ as follows.
First,
\begin{align}
	\mathcal{C}_{\hat{X}} &= \max_{e,a,a': a \neq a'} | \! \bra{e,a} \hat{X} \ket{e, a'} \! | 
	= \max_M \left\{ \frac{\Delta_M^2 \gamma_M^{\uparrow} + \Delta_{M+1}^2 \gamma_{M+1}^{\downarrow} }{ {}_N C_{N/2+M} } \right\}.
\end{align}
Because the most dominant coefficient is $\gamma_{1/2}^{\downarrow}$
in the superabsorption system due to the reservoir engineering,
the term $\mathcal{C}_{\hat{X}} $ is
approximately estimated as
\begin{align}
	\mathcal{C}_{\hat{X}} \simeq \frac{ \Delta_{1/2}^2 \gamma_{1/2}^{\downarrow} }{ {}_N C_{(N-1)/2} }.
\end{align}
Finally, the factor $\mathcal{C}_{l_1} ( \hat{\rho}_{\text{bd}} (t) )$ is evaluated as
\begin{align}
	\mathcal{C}_{l_1} ( \hat{\rho}_{\text{bd}}  (t) ) &= \sum_M p_M (t) \mathcal{C}_{l_1} ( \ket{M} \! \bra{M} ) \\
	&= \sum_M p_M (t) \sum_{(e,a) \neq (e',a') } | \! \braket{e,a | M } \! \braket{M | e', a' } \! | \\
	&= \sum_M \frac{ p_M (t) }{ {}_N C_{N/2 + M} } \sum_{a \neq a'}^{ {}_N C_{N/2 + M} }1 & &\left( \because \braket{e,a | M} = \frac{ \delta_{e,M} }{ \sqrt{ {}_N C_{N/2 + M }  } }   \right) \\
	 & = \sum_M \frac{ p_M (t) }{ {}_N C_{N/2 + M} } \times {}_N C_{N/2 + M} \left( {}_N C_{N/2 + M} - 1 \right).
\end{align}
Again, we assume the condition that most of the population confined in the 2-level system i.e. $p_{1/2} (t) + p_{-1/2} (t) \simeq 1$.
Then, $\mathcal{C}_{l_1} ( \hat{\rho}_{\text{bd}} (t) )$ can be further simplified as
\begin{align}
	\mathcal{C}_{l_1} ( \hat{\rho}_{\text{bd}}  (t) ) 
	&\simeq \sum_{M=-1/2, 1/2} p_M (t) \left( {}_N C_{N/2 + M} - 1 \right) \\
	&\simeq \sum_{M=-1/2, 1/2} p_M (t) {}_N C_{N/2 + M} & & \left( \because {}_N C_{N/2+M} \gg 1 \text{ for $M = - \frac{1}{2} \ \text{or} \ \frac{1}{2} $, and $N \gg 1$} \right) \\
	&=  {}_N C_{(N-1)/2}.
\end{align}
Therefore, the term $A_{\text{qm}} (t)$ and its scaling is evaluated as
\begin{align}
	A_{\text{qm}} (t) = \mathcal{C}_{\hat{X}} \mathcal{C}_{l_1} ( \hat{\rho}_{\text{bd}}  (t) )
	\simeq \frac{ \omega_A^2 \gamma_{1/2}^{\downarrow} }{ {}_N C_{ (N-1)/2 } } \times {}_N C_{ (N-1)/2 }
	= \omega_A^2 \gamma_{1/2}^{\downarrow}
	 = \Theta (N^2),
\end{align}
and this is much more dominant compared with
$A_{\text{cl}} (t) = \Theta \left( N^2 / 2^N \right)$.

\section{Comparison with the work by Tajima and Funo}
In this section, we clarify differences between our results
and those obtained by Tajima and Funo in Ref \cite{tajima2021superconducting}.
To this end, we first describe details of their model explicitly,
and then compare their heat engine model with ours.

\subsection{\textit{2$N_d$-state model} by Tajima and Funo}

\begin{figure}
	\begin{center}
		\includegraphics[clip,width=6cm,bb=0 0 1500 850]{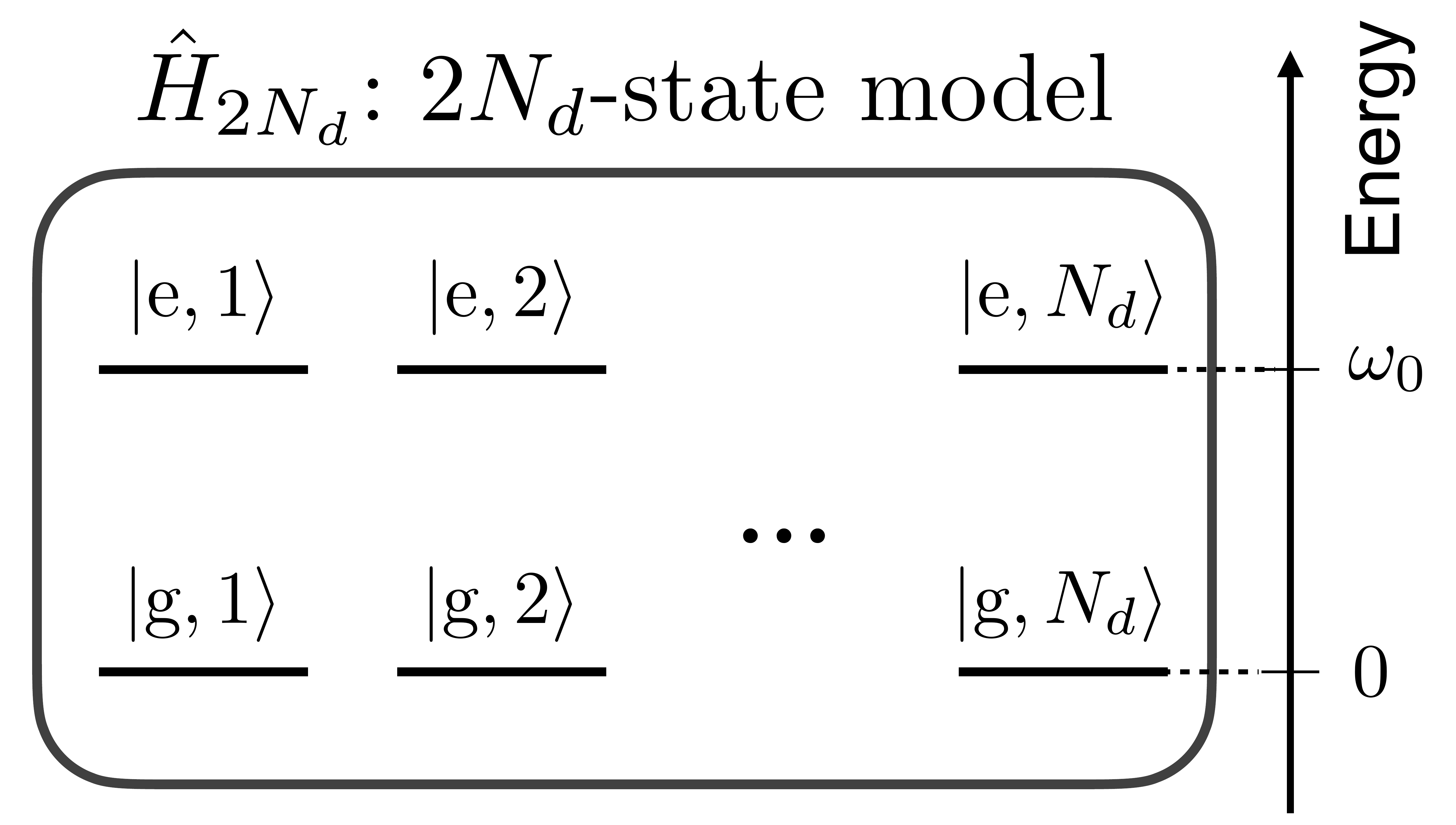}
		\caption{$2N_d$-state model introduced by Tajima and Funo
		}
		\label{fig:2Ndstate}
	\end{center}
\end{figure}

In Ref. \cite{tajima2021superconducting}, Tajima and Funo consider that a heat engine consists of a degenerate system whose Hamiltonian $\hat{H}_{2N_d} $ is given by
\begin{align}
	\hat{H}_{2N_d} = \sum_{j=1}^{N_d} \Bigl[ \omega_0 \ket{ \text{e}, j} \! \bra{ \text{e}, j } + 0 \ket{ \text{g}, j} \! \bra{ \text{g}, j } \Bigr]
	= \sum_{j=1}^{N_d} \omega_0 \ket{ \text{e}, j} \! \bra{ \text{e}, j }.
\end{align}
They call this system ``2$N_d$-state model". (cf. FIG. \ref{fig:2Ndstate}.)
This system has $2N_d$ energy eigenstates in total.
In this supplementary material, $N_d$ denotes the number of the excited states $\ket{\text{e}, j}$, which is the same as the number of the ground states $\ket{\text{g}, j}$.
As we can see from the explicit form of the Hamiltonian $\hat{H}_{2N_d}$,
the energy of the excited (ground) states
 is $\omega_0$ (0),
where we set $\hbar = 1$.

For the 2$N_d$-state model, Tajima and Funo assume the following
interaction Hamiltonian $\hat{H}_{\text{int}}^{2N_d}$ between the $2N_d$-state system and environment:
\begin{align}
	\hat{H}_{\text{int}}^{2N_d} = \sum_{j, j' = 1}^{N_d} \left( \ket{ \text{e}, j} \! \bra{ \text{g}, j'} \otimes \hat{B} + \ket{ \text{g}, j'} \! \bra{ \text{e}, j} \otimes \hat{B}^{\dagger}  \right),
	\label{eq:2Nd_interaction}
\end{align}
where $\hat{B}$ denotes an operator acting on the environment.
After taking the standard weak-coupling, Born-Markov, and rotating-wave approximations,
the GKSL master equation for a quantum state $\hat{\rho}_{2N_d} (t) $ at time $t$
is given as follows:
\begin{align}
	\frac{d \hat{\rho}_{2N_d} (t) }{dt} &= -i [ \hat{H}_{2N_d} , \hat{\rho}_{2N_d} (t) ] + \mathcal{D} [ \hat{\rho}_{2N_d} (t) ] , \\
	&\text{where} \ \mathcal{D} [ \hat{\rho} ] = \Gamma_{\downarrow} \left[ \hat{L} \hat{\rho} \hat{L}^{\dagger} - \frac{1}{2} \{ \hat{L}^{\dagger} \hat{L} , \hat{\rho} \} \right]
	+ \Gamma_{\uparrow} \left[ \hat{L}^{\dagger} \hat{\rho} \hat{L} - \frac{1}{2} \{ \hat{L} \hat{L}^{\dagger} , \hat{\rho} \} \right].
\end{align}
Specifically, they define a collective Lindblad operator $\hat{L}$ as
\begin{align}
	\hat{L} = \sum_{j = 1}^{N_d} \sum_{j' = 1}^{N_d} \ket{ \text{g}, j } \! \bra{ \text{e}, j'},
	\label{eq:jump_ope_2Ndmodel}
\end{align}
which generates a collective energy decay of the system.
Also, they assume that a detailed balance condition between the transition rates as $\Gamma_{\downarrow}/ \Gamma_{\uparrow} = e^{\beta \omega_0}$.

\subsection{Performance of $2N_d$-state model}
\begin{figure}
	\begin{center}
		\includegraphics[clip,width=12cm,bb=0 0 3250 2000]{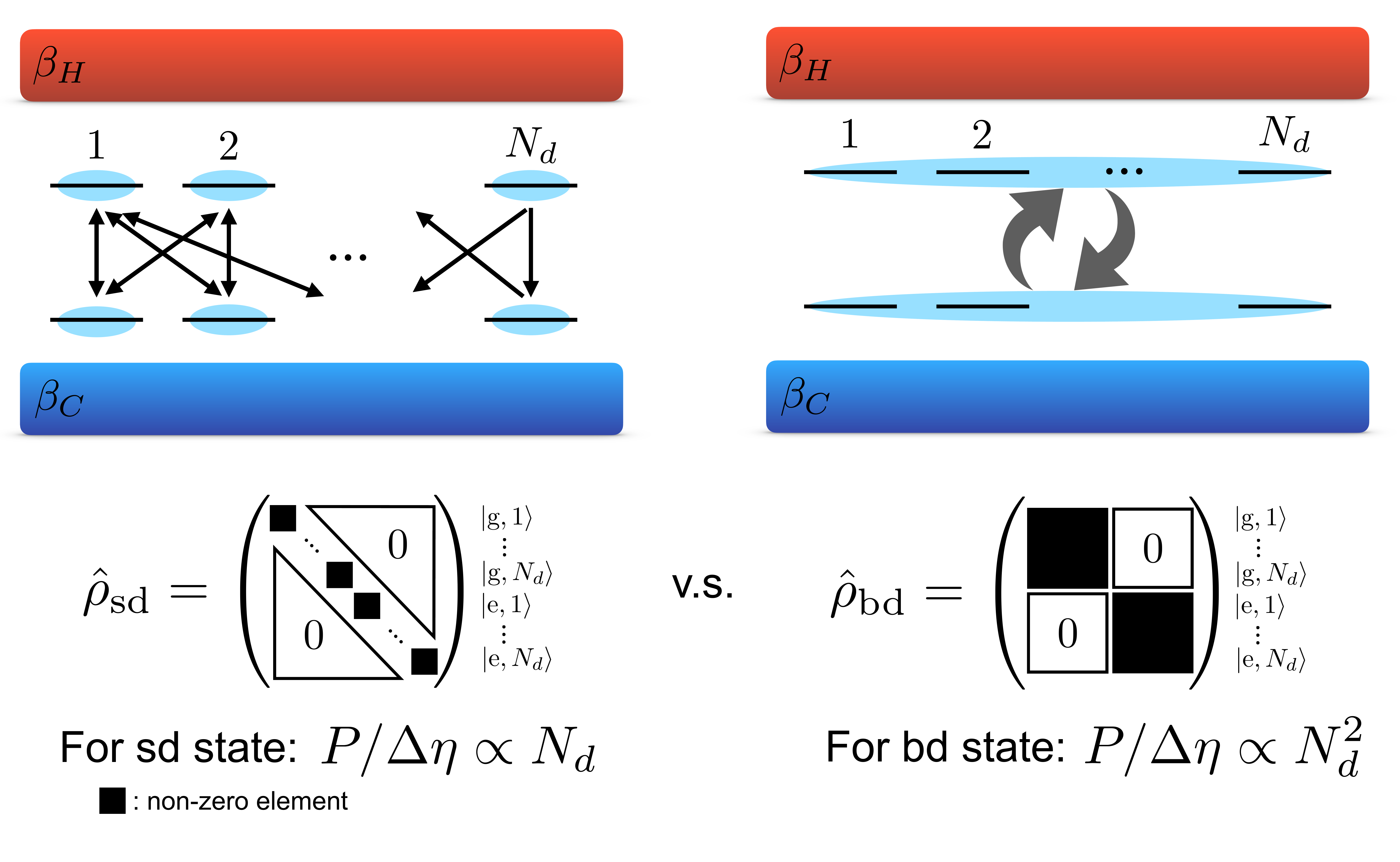}
		\caption{Scaling of the performance with
		the degeneracy $N_d$ of the 2$N_d$-state model discussed by Tajima and Funo}
		\label{fig:scal_2Nd}
	\end{center}
\end{figure}

Under these definitions introduced above, Tajima and Funo studied how the performance scales with
the degeneracy $N_d$.
For an sd state (defined as Eq. (\ref{eq:sd_state})) of the $2N_d$-state model,
they obtain the following scaling of performance:
\begin{align}
	\text{For an sd state:} \ \frac{P}{\Delta \eta} = \Theta (N_d).
	\label{eq:2Nd_sd_performance}
\end{align}
On the other hand, they also investigate a bd state (defined as Eq. (\ref{eq:bd_state})), and obtain
\begin{align}
	\text{For a bd state:} \ \frac{P}{\Delta \eta} = \Theta (N_d^2).
	\label{eq:2Nd_bd_performance}
\end{align}
From the above results Eq.~(\ref{eq:2Nd_sd_performance}) and Eq.~(\ref{eq:2Nd_bd_performance}),
they demonstrate a scaling enhancement of the heat engine performance.
Importantly, they firstly reveal that a quantum coherence among degenerate states can be utilized
to enhance the scaling with
the degeneracy $N_d$ (See FIG. \ref{fig:scal_2Nd}. for an abstract schematic of the scaling enhancement).

\subsection{Comparison with results by Tajima and Funo}

\begin{figure}
	\begin{center}
		\includegraphics[clip,width=12cm,bb=0 0 875 425]{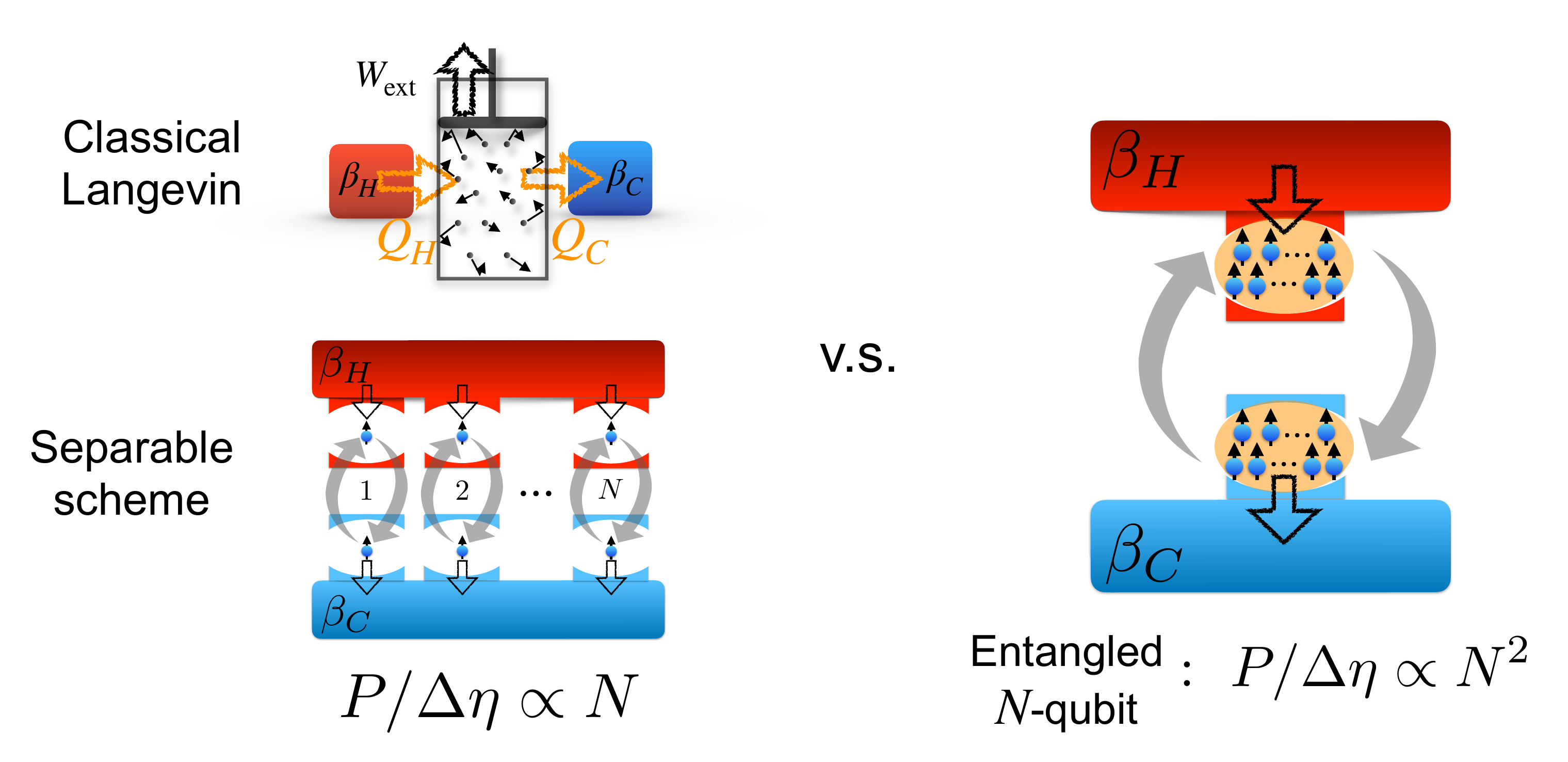}
		\caption{Scaling enhancement of the performance
		based on
		superabsorption.
		Unlike the 2$N_d$-state model by Tajima and Funo, we investigate how the performance scales not with the number of
		degeneracy $N_d$ but with
		the number of qubits $N$. As a result, we can compare our entangled scheme not only with the separable scheme, but also with the classical Langevin system.
		}
		\label{fig:scal_SAQHE}
	\end{center}
\end{figure}

Here, we compare the work done by Tajima and Funo with ours
(See Table \ref{tab:TFmodel} and Table \ref{tab:SKmodel}, respectively).
The parameter with
which Tajima and Funo study the scaling is the the number of
degeneracy $N_d$,
and they evaluate the scaling of performance for the two quantum states, 
the strictly diagonal (sd) state and the block-diagonal (bd) state
(cf. FIG. \ref{fig:scal_2Nd}).
Meanwhile, as discussed in detail in the main text, we study the scaling with
the number of qubits $N$,
and we compare the performance for the entangled $N$-qubit system (i.e. entangled scheme)
with that for a heat engine where we use $N$-qubit separable states
as the working media (i.e. separable scheme).
As a result, we obtain the performance $P/\Delta \eta = \Theta (N^2)$ for the entangled scheme,
and this exceeds the scaling for the separable scheme with $P/\Delta \eta = \Theta (N)$.
Also,
we can compare our result for the entangled scheme with
a classical heat engine performance $P/\Delta \eta = \Theta (N)$ for $N$ particles obeying a Langevin equation,
and we
clearly see the scaling advantage beyond such classical heat engines as well
(cf. FIG. \ref{fig:scal_SAQHE})

One might think that our model and results can be reproduced by an appropriate interpretation
of the $2N_d$-state model by Tajima and Funo.
However, it is not the case.
We have to notice that the Dicke states $\ket{1/2}$ and $\ket{-1/2}$
 (composing the E2LS in the main text)
are equal-weight superposition states of
roughly $2^N/\sqrt{N}$-degenerated states,
which is understood by a Stirling formula $n! \simeq \sqrt{2 \pi n} \left( \frac{n}{e} \right)^n$.
Thus, if we could naively adopt the result by Tajima and Funo,
the performance by the E2LS would scale as
$P/\Delta \eta = \Theta \left( \left( 2^N/\sqrt{N} \right)^2 \right) = \Theta (2^{2N}/N)$.
However, this does not coincide with our result $P/\Delta \eta = \Theta (N^2)$.
We can understand the apparently different two models and their results in a unified point of view as follows.


First, regarding the model of Tajima and Funo, the system-environment interaction is described as the interaction Hamiltonian in Eq.~(\ref{eq:2Nd_interaction}).
The jump operator $\hat{L}$ in Eq.~(\ref{eq:jump_ope_2Ndmodel}) is exactly the same as the system operator of the interaction Hamiltonian, where information of the environment is embedded only in the correlation function $\Gamma_{\downarrow (\uparrow)}$.
In their enhanced heat engine, they consider a subspace that is spanned by the following two equal-weight superposition states ($\ket{\text{e}, +}$ and $\ket{\text{g}, +}$) of degenerate quantum states:
\begin{align}
	\ket{\text{e}, +} &= \frac{1}{\sqrt{N_d}} \sum_{j = 1}^{N_d} \ket{\text{e}, j}, &
	\ket{\text{g}, +} &= \frac{1}{\sqrt{N_d}} \sum_{j = 1}^{N_d} \ket{\text{g}, j}.
\end{align}
Then, the transition rate between these two states under the GKSL master equation is given by the square of the absolute value of the matrix element $\bra{ \text{g}, + } \hat{L} \ket{ \text{e}, +}$.
This element explicitly reads
\begin{align}
	\bra{ \text{g}, + } \hat{L} \ket{ \text{e}, +} 
	&= \left[ \frac{1}{ \sqrt{N_d} }  \Bigl( \bra{ \text{g}, 1 } + \cdots + \bra{ \text{g}, N_d }  \Bigr) \right]
	\sum_{k, k' = 1}^{N_d} \ket{ \text{g}, k } \! \bra{ \text{e}, k'} 
	\left[ \frac{1}{ \sqrt{N_d} }  \Bigl( \ket{ \text{e}, 1 } + \cdots + \ket{ \text{e}, N_d }  \Bigr) \right]
	\\
	&= \frac{1}{N_d} \Bigl( \bra{ \text{g}, 1 } + \cdots + \bra{ \text{g}, N_d } \Bigr) 
	\sum_{k, k' = 1}^{N_d} \ket{ \text{g}, k } \! \bra{ \text{e}, k'} 
	\Bigl( \ket{ \text{e}, 1 } + \cdots +  \ket{ \text{e}, N_d } \Bigr) 
\end{align}
Here, by the definition of the interaction Hamiltonian, the jump operator $\hat{L}$ connects one single excited state $\ket{ \text{e}, j'}$ to every single ground state $\ket{ \text{g}, j}$,
which one can see from $\bra{ \text{g}, j} \hat{L} \ket{ \text{e}, j'} = 1$ for all $(j, j')$.
This relation reflects the specific feature of the model of Tajima and Funo;
in their model, the number of possible relaxation (excitation) processes from one single excited (ground) state to ground (excited) states, which we call the {\it connectivity}, is given by $N_d$.
Because we obtain $N_d$ for one single excited state $\ket{ \text{e}, j'}$ from the right parentheses, then, we obtain
\begin{align}
	\bra{ \text{g}, + } \hat{L} \ket{ \text{e}, +} 
	&= \frac{1}{N_d} \Bigl( \underbrace{N_d}_{ \ket{\text{e}, 1} } + \underbrace{N_d}_{ \ket{\text{e}, 2} } + \cdots + \underbrace{N_d}_{ \ket{\text{e}, N_d} } \Bigr)
	= \frac{1}{N_d} \times N_d \times N_d
	= N_d.
\end{align}
Thus, the square of the absolute value of this matrix element provides us the transition rate between $\ket{ \text{e}, +} $ and $\ket{ \text{g}, +}$, which is proportional to $N_d^2$.
This leads to the scaling $\Theta (N_d^2)$ of the speed of dynamics, and the enhanced heat engine performance $P / \Delta \eta = \Theta (N_d^2)$.

On the other hand, in our system, the subspace under consideration is spanned by the two Dicke states $\ket{1/2}$ and $\ket{-1/2}$.
The Dicke state $\ket{1/2}$ ($\ket{-1/2}$) is an equal-weight superposition state of all the $N$-qubit computational bases having $ \frac{N+1}{2} $ $\left(  \frac{N-1}{2} \right)$ qubits in the excited state $\ket{ \text{e} }$ and the other $ \frac{N-1}{2} $ $\left(  \frac{N+1}{2} \right)$ qubits in the ground state $\ket{ \text{g} }$ (cf. $N$: odd):
\begin{align}
	\left|  \frac{1}{2} \right\rangle &= \frac{1}{ \sqrt{ {}_N C_{(N+1)/2 } } }
	\sum_{ \text{all configuration} } | \{ \underbrace{ \text{e} \cdots \text{e} }_{ \frac{N+1}{2} } \underbrace{ \text{g} \cdots \text{g} }_{ \frac{N-1}{2} } \} \rangle , &
	\left| - \frac{1}{2} \right\rangle &= \frac{1}{ \sqrt{ {}_N C_{(N+1)/2 } } }
	\sum_{ \text{all configuration} } | \{ \underbrace{ \text{e} \cdots \text{e} }_{ \frac{N-1}{2} } \underbrace{ \text{g} \cdots \text{g} }_{ \frac{N+1}{2} } \} \rangle .
\end{align}
In our system, the system-environment interaction Hamiltonian contains the operator $\hat{J}_{\pm} = \sum_{i=1}^{N} \hat{\sigma}_{\pm}^{(i)}$ of the system.
Thus, the relevant matrix element is $\bra{-1/2} \hat{J}_- \ket{1/2}$,
which is explicitly given by
\begin{align}
	\bra{-1/2} \hat{J}_- \ket{1/2} &= 
	\left( \frac{1}{ \sqrt{ {}_N C_{(N+1)/2 } } } \sum_{ \text{a.c.} } \bra{ \{ \underbrace{ \text{e} \cdots \text{e} }_{ \frac{N-1}{2} } \underbrace{ \text{g} \cdots \text{g} }_{ \frac{N+1}{2} } \} } \right)
	\sum_{i=1}^{N} \hat{\sigma}_{ - }^{(i)}
	\left( \frac{1}{ \sqrt{ {}_N C_{(N+1)/2 } } } \sum_{ \text{a.c.} } \ket{ \{ \underbrace{ \text{e} \cdots \text{e} }_{ \frac{N+1}{2} } \underbrace{ \text{g} \cdots \text{g} }_{ \frac{N-1}{2} } \} } \right) \\
	&=  \frac{1}{  {}_N C_{(N+1)/2 } } 
	\left( \sum_{ \text{a.c.} } \bra{ \{ \underbrace{ \text{e} \cdots \text{e} }_{ \frac{N-1}{2} } \underbrace{ \text{g} \cdots \text{g} }_{ \frac{N+1}{2} } \} } \right)
	\sum_{i=1}^{N} \hat{\sigma}_{ - }^{(i)}
	\left( \sum_{ \text{a.c.} } \ket{ \{ \underbrace{ \text{e} \cdots \text{e} }_{ \frac{N+1}{2} } \underbrace{ \text{g} \cdots \text{g} }_{ \frac{N-1}{2} } \} } \right).
	\label{eq:explct_form_matelSA}
\end{align}
To get an intuitive idea of this calculation, we write down the transition from the specific computational basis $\ket{ \text{eeegg} }$ for $N = 5$:
\begin{align}
	\ket{ \text{eeegg} } \xrightarrow{ \hat{J}_{-} }  \ket{ \text{eeggg} } + \ket{ \text{egegg} } + \ket{ \text{geegg} }.
\end{align}
We have a superposition of three computational basis states.
The number ``3" of the flipped states comes from the number of the excited states (in the initial state $\ket{ \text{eeegg} }$), which we can flip by the operator $\hat{J}_-$.
Going back to Eq.~(\ref{eq:explct_form_matelSA}) and taking one single computational basis $\ket{ \{ \underbrace{ \text{e} \cdots \text{e} }_{ \frac{N+1}{2} } \underbrace{ \text{g} \cdots \text{g} }_{ \frac{N-1}{2} } \} }$ in the right parentheses,
we have $ \frac{N+1}{2}$ excited states $\ket{ \text{e} }$ to flip to the ground states $\ket{ \text{g} }$ and obtain nonzero matrix elements $\left( \text{cf.} \ \frac{N+1}{2} = 3 \ \text{for} \ N=5 \right)$.
Then, we can calculate the matrix element $\bra{-1/2} \hat{J}_- \ket{1/2}$ as
\begin{align}
	\bra{-1/2} \hat{J}_- \ket{1/2} 
	&= \frac{1}{  {}_N C_{(N+1)/2 } }
	\Biggl( \underbrace{ \frac{N+1}{2}   + \frac{N+1}{2}  + \cdots + \frac{N+1}{2} }_{ {}_N C_{(N+1)/2 } \ \text{terms} } \Biggr) 
	= \frac{1}{  {}_N C_{(N+1)/2 } } \times  {}_N C_{(N+1)/2 } \times \frac{N+1}{2} \\
	&= \frac{N+1}{2}.
\end{align}
Thus, the square of the absolute value of this matrix element leads to the transition rate between $\ket{1/2} $ and $\ket{-1/2}$, which is proportional to $a_{1/2} = \frac{1}{4} (N+1)^2$.

Combining the results for the model of Tajima and Funo and ours, 
we can see the similarity between these two, which 
is illustrated in a simple form as follows:
\begin{align}
	(\text{matrix element}) = \frac{1}{ (\text{
	normalization factor
	})^2 } \times  \text{(number of degeneracy)} \times ( \text{connectivity} ) = ( \text{connectivity} ),
\end{align}
where we use the fact that the square of the normalization factor of an equally-weight superposition state is equivalent
to the number of degeneracy.
Therefore, we can conclude that the crucial factor leading to the enhanced scaling of performance is not the degeneracy in general; rather the connectivity remains after the calculation of the matrix element, and its square results in the scaling of the transition rates, or equivalently, the scaling of power.
Incidentally, this importance of the connectivity was not so clear in the paper by Tajima and Funo,
because, in their model, the number of degeneracy coincides with the connectivity, and we cannot easily distinguish these two independent contributions to the scaling of the matrix element.
Taking this unified point of view, we can understand both the results of Tajima and Funo and ours, and we can get an insight into how to enhance the scaling of heat engine performance in general settings.
Specifically, in order to obtain the enhanced scaling for a quantum heat engine,
we have to implement a system-environment interaction that induces a large number of possible transitions between one degenerate subspace and another.

\begin{table}[H]
\caption{Tajima Funo model}
\label{tab:TFmodel}
\centering
\begin{tabular}{l |c}
\hline
Parameter of scaling & $N_d$: number of degeneracy \\
Scheme with sd state & $P/\Delta \eta = \Theta (N_d)$ \\
Scheme with bd state &  $P/\Delta \eta = \Theta (N^2_d)$ \\
Comparison & sd state v.s. bd state
\end{tabular}
\end{table}

\begin{table}[H]
\caption{Our model}
\label{tab:SKmodel}
\centering
\begin{tabular}{l |c }
\hline
Parameter of scaling &$N$: number of qubits \\
Separable scheme & $P/\Delta \eta = \Theta (N)$ \\
Classical counterpart & Langevin system: $P/\Delta \eta = \Theta (N)$ \\
Scheme with entangled states &  $P/\Delta \eta = \Theta (N^2)$ \\
Comparison & separable scheme (or Langevin sys.) v.s. entangled scheme
\end{tabular}
\end{table}

\section{Experimental realization}
Here, we explain a possible physical realization of our scheme using
superconducting flux qubits.
Superconducting qubits are artificial atoms with significant design freedoms \cite{clarke2008superconducting}.
Especially, a superconducting flux qubit is composed of a superconducting loop interrupted by Josephson junctions.
When we change the parameters of the circuit, we can design suitable properties of the flux qubits.
For example, we can tune the qubit frequency 
and coupling strength 
by using magnetic flux through the SQUID
 embedded in the flux qubit \cite{paauw2009tuning,zhu2010coherent,harris2010experimental,2harris2010experimental}.
 Moreover, the flux qubit can be strongly coupled with a microwave cavity \cite{abdumalikov2008vacuum,yamamoto2014superconducting,lindstrom2007circuit,johansson2006vacuum,chiorescu2004coherent}. There are experimental demonstrations that an ensemble of the flux qubits is coupled with the microwave cavity \cite{macha2014implementation,kakuyanagi2016observation}.
 A long coherence time of around 90 ms
 is achieved \cite{yan2016flux,bylander2011noise,abdurakhimov2019long}. 
Also, a technique to realize a coupling between the superconducting qubits via airbridged microwave cavity is reported
 \cite{mukai2020pseudo}, and this is useful to couple distant qubits.
  The cavity lifetime can be as long as a milli second \cite{reagor2016quantum}, and we can  decrease the cavity lifetime depending on the purpose
 \cite{sevriuk2019fast}.
  The coupling strength between the flux qubit and cavity
can be as large as a few GHz \cite{yoshihara2017superconducting}
 while the typical frequency of the flux qubit is also a few GHz. Also, the coupling between the flux qubit can be as large as a few GHz \cite{harris2010experimental,2harris2010experimental}.
 These properties are prerequisite to realize our scheme.
 
 In our proposed scheme, we consider 2$N$ flux qubits.
We assume that $N$ flux qubits are coupled with an LC resonator (which we call a cavity A)
 that is coupled with a low-temperature thermal bath, while the other flux qubits are coupled with a different LC resonator (which we call a cavity B) that is coupled with a high-temperature thermal bath \cite{johansson2006vacuum}.
 We can couple the $N$ flux qubits with the other $N$ flux qubits via a transmission line resonator \cite{lindstrom2007circuit,macha2014implementation}.
 Since the transmission line can be as large as a few cm, 
 the temperature of the bath coupled with the cavity A could be different from that coupled with the cavity B.
 We can perform SWAP gates between the flux qubits by using the transmission line resonator, and this allows us to couple quantum states either a high temperature bath or low temperature bath
 \cite{blais2004cavity}.

 Our scheme can be implemented with the flux qubits as follows.
 First, we thermalize the $N$ flux qubits coupled with the cavity A. Second, we change the frequency of the flux qubits at the cavity A, and swap the quantum state of the qubits coupled with the cavity A to the one coupled with the cavity B.
 Third, we thermalize the $N$ qubits at the cavity B.
 Fourth, we change the frequency of the flux qubits at the cavity B, and swap the quantum state of the qubits coupled with the cavity B to that coupled with the cavity A. Finally, we repeat these four steps.
 
In the numerical calculations shown in FIG.2 in the main text,
we assume that the frequency of the flux qubits is $\omega _A^H/2\pi=1$ GHz.
Then, the coupling strength $\Omega / 2\pi$ between the qubits is equal to 31 MHz.
By setting the coupling strength between the transmission line resonator and the flux qubits to be around 100 MHz,
we can perform the swap gates within a few nano seconds.
During the thermalization processes, the cavity-qubit coupling $g/2 \pi$ is around $10^{-5} \times \omega_A^H/ 2 \pi = $10 kHz,
and the thermalization period $\tau_H$ and $\tau_C$ is around 10 nano seconds when $N=9$,
while the coherence time of the flux qubits can be as long as 90 ms.
Thus, we can ignore the decoherence of the flux qubits except that induced by the cavity A and B.
The temperature inside the dilution refrigerator can be as small as 10 mK,
and the temperature $T_H = (k_B \beta_H)^{-1}$ and $T_C = (k_B \beta_C)^{-1}$ we assume
are respectively given by 20 mK and 10 mK.

\bibliographystyle{apsrev4-1}
\bibliography{SAQHE_suppl}